%% file: main.tex
\documentclass[usenatbib]{mnras}

\usepackage{newtxtext,newtxmath}

\usepackage[T1]{fontenc}

\DeclareRobustCommand{\VAN}[3]{#2}
\let\VANthebibliography\thebibliography
\def\thebibliography{\DeclareRobustCommand{\VAN}[3]{##3}\VANthebibliography}

\usepackage[utf8]{inputenc} 
\usepackage[T1]{fontenc}    
\usepackage{hyperref}       
\usepackage{url}            
\usepackage{booktabs}       
\usepackage{amsfonts}       
\usepackage{nicefrac}       
\usepackage{microtype}      
\usepackage[table,x11names]{xcolor}         
\usepackage{graphicx}
\usepackage{caption}
\usepackage{subcaption}
\usepackage{color,soul}
\usepackage{placeins}
\graphicspath{ {./img/} }
\hypersetup{urlcolor=cyan, 
            linkcolor=blue,
            colorlinks=true}
\usepackage{multirow}

\usepackage{tabularx}
\usepackage{adjustbox}
\usepackage{array}
\usepackage{hhline}

\newcolumntype{R}[2]{%
    >{\adjustbox{angle=#1,lap=\width-(#2)}\bgroup}%
    l%
    <{\egroup}%
}
\newcommand*\rot{\multicolumn{1}{R{45}{1em}}}

\usepackage{breakcites}

\author[]{
Rabea Sennlaub$^{\ast}$,$^{1}$
Martin Hofmann$^{\ast}$,$^{1}$\\
Mike Hankey,$^{2}$
Mario Ennes,$^{3}$
Thomas Müller,$^{3}$
Peter Kroll,$^{3}$
Patrick Mäder$^{1}$
\\
$^{1}$Department of Computer Science, Technical University Ilmenau, Ehrenbergstraße 29, Ilmenau, Germany\\
$^{2}$Operations, American Meteor Society, 54 Westview Crescent, Geneseo , NY 14454, USA\\
$^{3}$Wide-Area-Sky-Monitoring, Sonneberg Observatory, Sternwarte Straße 32, Sonneberg, Germany\\
$^{\ast}$These authors contributed equally.\\
\textbf{This is a pre-copyedited, author-produced PDF of an article accepted for publication in Monthly Notices of the Royal Astronomical Society following peer review.}\\ \textbf{The version of record is available online at: \url{http://doi.org/10.1093/mnras/stac1948}}.
}

\title{Object classification on video data of meteors and meteor-like phenomena: algorithm and data}

\pubyear{2022}

\begin{document}

\maketitle

\input{content/abstract}

\begin{keywords}
astronomical data bases: miscellaneous -- methods: data analysis -- techniques: image processing -- meteors
\end{keywords}

\input{content/introduction}
\input{content/dataset}

\input{content/applications}

\input{content/experiments}
\input{content/limitations}
\input{content/conclusions}

\input{content/ack}

\input{content/ref}

\appendix
\input{content/datasheet}

\end{document}

%% file: content/abstract.tex

\begin{abstract}

Every moment, countless meteoroids enter our atmosphere unseen. The detection and measurement of meteors offer the unique opportunity to gain insights into the composition of our solar systems' celestial bodies. Researchers, therefore, carry out a wide-area-sky-monitoring to secure 360-degree video material, saving every single entry of a meteor. Existing machine intelligence cannot accurately recognize events of meteors intersecting the earth's atmosphere due to a lack of high-quality training data publicly available. This work presents four reusable open source solutions for researchers trained on data we collected due to the lack of available labeled high-quality training data. We refer to the proposed dataset as the NightSkyUCP dataset, consisting of a balanced set of 10,000 meteor- and 10,000 non-meteor-events. Our solutions apply various machine learning techniques, namely classification, feature learning, anomaly detection, and extrapolation. For the  classification task, a mean accuracy of 99.1\% is achieved. The code and data are made public at figshare with DOI \hyperlink{https://doi.org/10.6084/m9.figshare.16451625}{10.6084/m9.figshare.16451625}.

\end{abstract}

%% file: content/introduction.tex

\section{Introduction}
\begin{figure*}
    \begin{subfigure}[b]{0.49\textwidth}
		\centering
		\includegraphics[width=\textwidth]{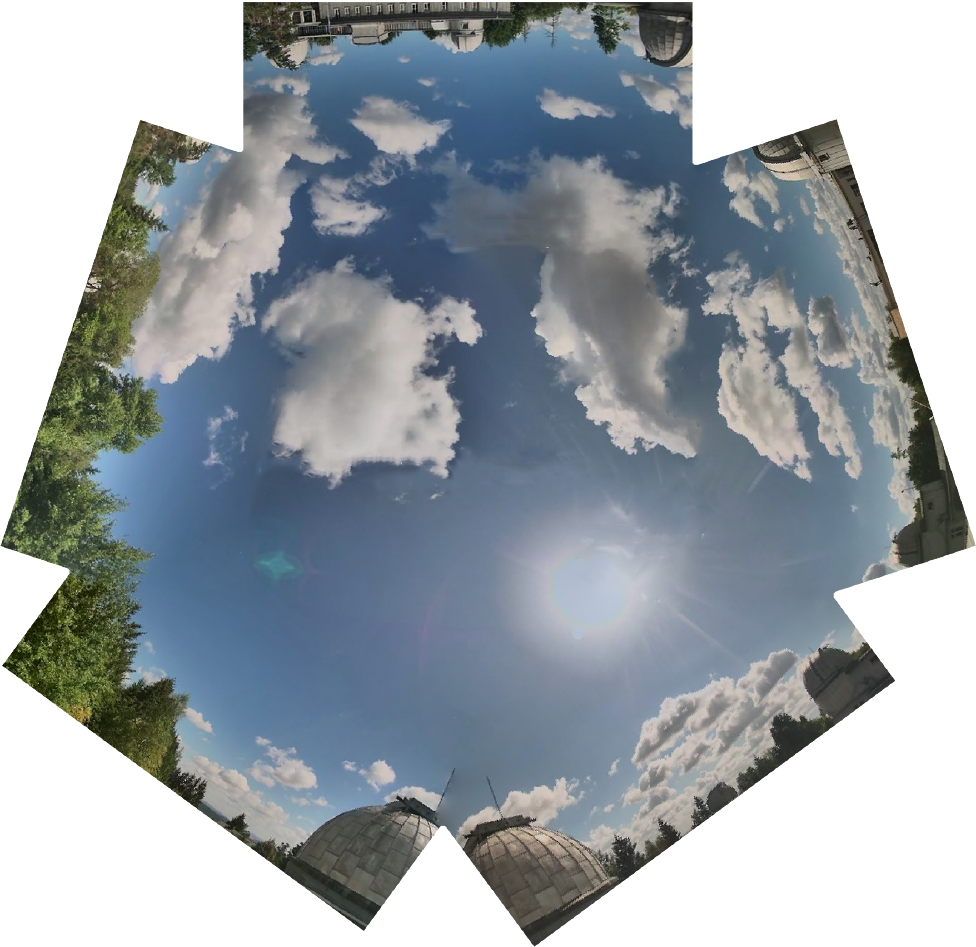}
		\caption{Sonneberg observatory by day.}
	\end{subfigure}
	\begin{subfigure}[b]{0.49\textwidth}
		\centering
		\includegraphics[width=\textwidth]{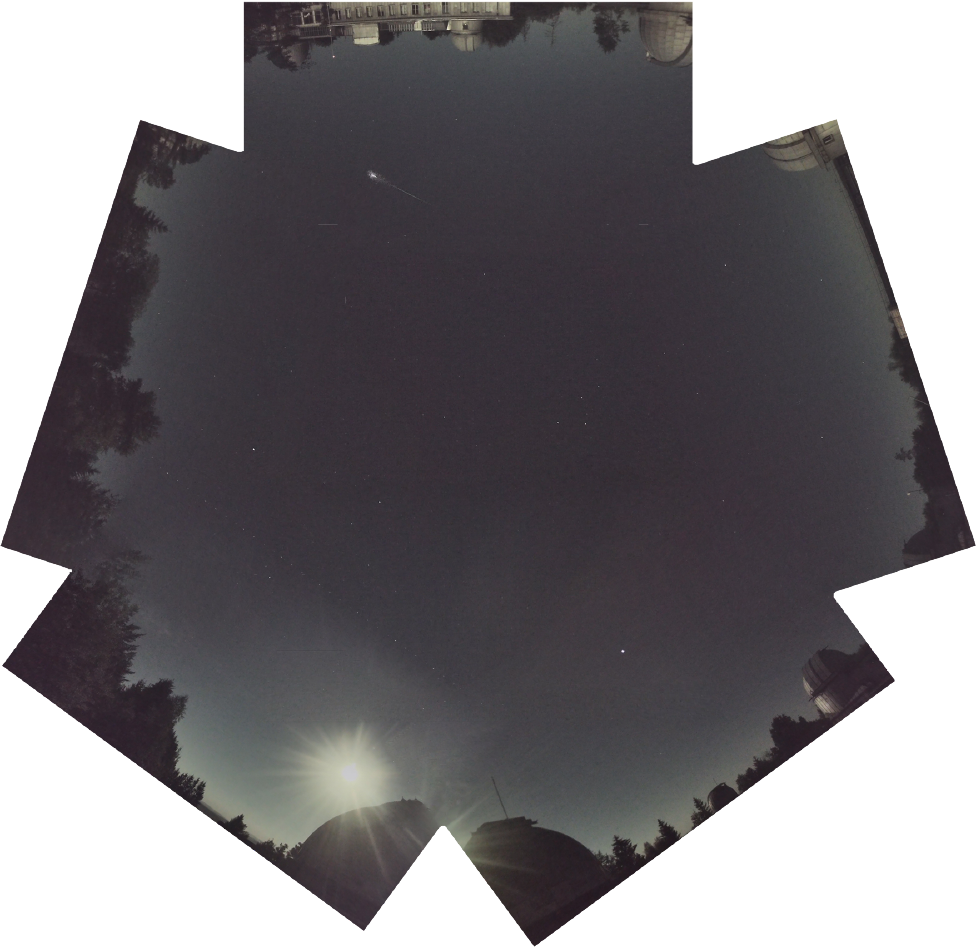}
		\caption{Sonneberg observatory by night with perseid.}
	\end{subfigure}
	\caption{Stiched raw video material of the camera arrangement of the AllSky7 system using an observation at the Sonneberg station. The source data is available online (cp. \citep{AllSkyPub})}
	\label{fig:allsky7_cameraposition}
\end{figure*}

Meteors have always been a source of fascination for people. In the past, meteors were divine signs or harbingers of misfortune, like wars, plague, and bad harvests. This view changed at the end of the 18th century with a publication of Ernst Florens Friedrich Chladni, who set up in his book the revolutionary and controversial thesis that the meteorites found on earth have their origin in space \cite{Chladni1794}. Howard and de Bournon published scientific data on meteorites for the first time only a few years later \cite{Buehler1988}. Meteoroids are the solar system's small and smallest body fragments we can observe falling into the earth's atmosphere. If these enter the earth's atmosphere, luminous phenomena appear, which are called meteors. If meteoroids or parts of meteoroids reach the earth's surface, they are called meteorites. The detection and measurement of meteors offer many possibilities to gain scientific knowledge. For example, meteor observations can be used to determine an orbit, which provides insight into the origin of these small bodies \cite{Janches2020, Ferus2020}. Furthermore, the observed sequence of luminosity and possibly spectroscopy allows conclusions about the composition of the material and its size \cite{Ferus2020}. In an effort to capture every single meteor entry, researchers carry out a wide-area-sky-monitoring. Since early 2019, there has been an independent network of digital stations under construction and continuous expansion called \textit{AllSky7 Fireball Network Germany}, which is mainly operated by volunteers \cite{AllSky7}. Currently, the network consists of over 40 stations across the US and central Europe. Each station has seven cameras arranged in a hemisphere, covering the entire sky with overlap (see Figure~\ref{fig:allsky7_cameraposition}). Generic calibrated board cameras with IMX 291 Sony Starvis CMOS\footnote{Complementary metal-oxide-semiconductor} chips are used here because they are small, inexpensive, and suitable for low-light applications \cite{Hankey2020}. 
Camera operators and specialists classify thousands of false positive samples by hand due to the lack of an accurate pre-classification model at AllSky7. Accordingly, our task was to develop machine learning algorithms and a data-collection process that yields high-quality data. Our intensive search did not yield any publicly available data helping us to further enhance the pre-classification model at AllSky7. Therefore, scientists and voluntary enthusiasts decided to collect those high-quality data to train novel machine learning algorithms, reducing the immense workload.

We collected and proposed the NightSkyUCP\footnote{NightSky Unidentified Celestial Phenomena} dataset with the idea in mind to bring the machine learning algorithms and the data to the people who need them and to eventually reduce the classification efforts of camera operators. Table~\ref{tab:other_datasets} gives an overview of previously used datasets in the field. Whether a classifier can effectively detect non-meteor events by learning what meteors look like is an open question that we shed light on using our collected data. However, public datasets consist of only one class. Our dataset closes this gap and enables researchers and enthusiasts to benchmark approaches against another and develop new methods. The NightSkyUCP dataset consists of equally distributed meteor and non-meteor samples. Although meteor events are relatively rare events observable in the sky, they are usually easily recognizable. Therefore, we focus on the fewer non-meteor events that have been confused by our detection and classification pipeline and are classified manually by our camera operators; These samples are considered 'hard' samples.

After all, we tackled the challenge of creating a model that solves the pre-categorization task to give volunteers and experts the time to collect more valuable data. We refer to this task as classification. A problem with meteor videos is that objects in the foreground sometimes obstruct the trajectory, and so, several videos of one single event are created. The prediction of events curvature enables the matching of events with obstructed trajectories.  Matching events by their anticipated curvature produces more conclusive i.e., less redundant events and, finally, higher-quality samples. We refer to this task as extrapolation. Due to the small number of labeled sub-class samples, we evaluate their conclusiveness qualitatively. We, therefore, visualize their extracted hidden representations, referring to this task as feature learning. Finally, we evaluated the quality of an algorithm trained only on meteor data to determine if the already available videos of meteor events would have been enough to create a high-quality classifier. Since we model the data distribution of the meteor class and compare it to the data distribution of the non-meteor class, we refer to this task as anomaly detection, i.e., we try to detect samples that do not fit the data distribution (anomalies).

These methods show exciting results and novel approaches, but few provided open data and the code. Thus, we decided to make our algorithms and data public and publicly available. Accordingly, we integrate the mentioned techniques, such as pre-trained feature extractors and recurrent neural networks, and evaluated various classifiers, recurrent units, and input data variations such as single images or stacks representing video sequences.

Methods developed based on our dataset could be used in the future not only for meteors but also for other transient celestial meteor-like events.
One exciting example is space debris research which plays an increasing role in connection with mega-constellations of satellites. 
The reentering satellite and rocket debris may show very similar characteristics as meteors. 
It might be possible to differentiate between them by the duration of the glow, the course of brightness, and the absolute brightness. 

Table~\ref{tab:other_datasets_video} compares our dataset with two of the most popular general video databases.
It is noticeable that while having fewer samples, overall, NightSkyUCP disposes more samples per class and is balanced.
Another advantage of NightSkyUCP for the Machine Learning community is that videos of the two classes, meteor, and non-meteor, are very similar and often differ only in minor details, such as curvature, direction, color variation, or fading. Comparing meteor videos to videos of planes, one notices only bright objects with dark backgrounds in the first place. Only experts can distinguish such samples, especially if the sequences are short. Therefore it can be an exciting challenge not only for classification but also for anomaly detection.

With this work, we provide the algorithms needed to start classifying sky phenomena along with a publicly available dataset of sky phenomena consisting of both meteors and non-meteors with a total of $20,000$ samples. At least three different experts evaluated every sample to ensure a high quality of the dataset. Two more experts additionally checked 297 non-meteor and 277 meteor samples to ensure they were correctly labeled. These samples were only accepted if all five experts came to a conclusive decision. We provide videos of the events, sum-images, stacks of crops, and motion data to accommodate a wide range of tasks. NightSkyUCP is available to the reviewers at \hyperlink{https://figshare.com/s/0c88dc958350e32020d7}{figshare} and will be published under a \hyperlink{https://creativecommons.org/licenses/by-nc-sa/4.0/}{CC BY-NC-SA} license once accepted. A datasheet \cite{gebru2018datasheets} summarizing NightSkyUCP along with detailed documentation can be found in the Appendix~A.

The NightSkyUCP can be used for a variety of tasks, namely classification, feature learning, anomaly detection and extrapolation. 
Solving of these tasks can help to improve meteor detection, classification of different transient celestial events and provide deeper insight in the nature of the data.
To explore the possibilities we apply several advanced machine learning methods.

The remaining sections are organized as follows: Section \ref{sec:dataset} describes the NightSkyUCP dataset, including data collection, preprocessing and analysis. Section \ref{sec:experiments} describes the experiments for the tasks classification, feature learning, anomaly detection and extrapolation. In section \ref{sec:limitations} the limitations of the NightSkyUCP dataset are assessed, followed by a conclusion in section \ref{sec:conclusion}.

\begin{table*}
    \centering
    \caption{Previously studied and partly publicly available meteor datasets}
    \begin{tabular}{lp{.2\textwidth}rlllp{.2\textwidth}l}
        \hline
        \textbf{Dataset} & \rot{\textbf{Ref.}} & \rot{\textbf{\#Samples}} & \rot{\textbf{Classes}} & \rot{\textbf{Videos}} & \rot{\textbf{Sum-images}} & \textbf{Metadata} & \rot{\textbf{Availability}} \\ \hline
        AstDyS-2 & \cite{AstDyS-2} & 931,796 & single& no & no & orbital data & public \\ \hline
        NASA Fireballs & \cite{NasaAllSky} & 98 & single& yes & yes & Location, light curve, orbital data & public \\ \hline
        SNM20xxx & \cite{Sonota} & 346,041 & single& no & no & radiant maps, ground maps, orbital data & public \\ \hline
        Video MeteorDB meta & \cite{MeteorDB} & 3,971,618 & single& no & no & coordinates, apparent velocity, brightness, shower membership & public \\
        \hline
        EDMOND 5 v.04 & \cite{kornos2014edmond} & 317,380 & single & no & no & orbital data, atmospheric parameters & public \\ \hline
        CMN Orbit Catalogues & \cite{korlevic2013croatian} & 41.634 & single & no & no & orbital data & public \\ \hline
        CAMS Meteoroid Orbit Database v3.0 & \cite{jenniskens2018survey} & 471,582 & single & no & no & orbital data & public \\ \hline
        Global meteor network dataset & \cite{vida2021global} & >220,000 & single & no & no & orbital data & public \\ \hline
        Video MeteorDB img & \cite{MeteorDB} & 3,971,618 & single& yes & yes & position, brightness per frame & private \\ \hline
        CAMS video ML subset & \cite{Gural2019} & $\approx$200,000 & multi & yes & yes & track measurements & private \\
        \hline\hline
        \textbf{NightSkyUCP (ours)} &  & 20,000 & multi& yes & yes & stacks of crops and movement & public \\ \hline
    \end{tabular}
    \label{tab:other_datasets}
\end{table*}

\begin{table*}
    \centering
    \caption{Comparison of popular video datasets with NightSkyUCP}
    \begin{tabular}{llrlll}
        \hline
        \textbf{Dataset} & \rot{\textbf{Ref.}} & \rot{\textbf{\#Samples}} & \rot{\textbf{\#Classes}} & \rot{\textbf{$\frac{\#Samples}{Class}$}} & \rot{\textbf{balanced}} \\ \hline
        YouTube-8M & \cite{youTube8M} & 6.1Million & 3862 & avg. of 3552 & no\\ \hline
        Kinetics 700 & \cite{carreira2019short} & 650.317 & 700 & min. 600 & no\\ \hline\hline
        \textbf{NightSkyUCP (ours)} &  & 20.000 & 2 & 10.000 & yes\\ \hline 
    \end{tabular}
    \label{tab:other_datasets_video}
\end{table*}

%% file: content/dataset.tex

\section{The NightSkyUAP Dataset}\label{sec:dataset}
The NightSkyUAP dataset provides a publicly available source of both meteor and non-meteor video data for classification, motion prediction, unsupervised clustering, and anomaly detection. The dataset consists of 20,000 H.264 encoded color videos of phenomena inside MPEG-4 container files, absolute and relative positions of the phenomenon in the video and class information in a CSV file, corrected colored sum-images of the video files as jpeg image-files, and sequences of 32x32 pixel crops as a NumPy array stack, stored in python pickle format.

\subsection{Data Collection and Annotation}
The cameras of the Allsky7 network run continuously and record two data streams in parallel, one in HD ($1920\times1080$ pixels) and one in SD ($704\times576$ pixels), both at 25 frames per second.
The image sequences are H.264 encoded into MPEG files, each one minute long and held for up to two days. Image analysis software runs at each station to detect interesting events that could be meteors in the SD video clips. To obtain only information potentially relevant for classification, the videos are spatially and temporally cropped to these events. Both the detection and the cropping are performed using the AllSky7 software, which is accessible at \cite{amsCams}.

\begin{figure*}
    \centering
    \includegraphics[trim=0 0 5 0, clip,width=1.00\textwidth]{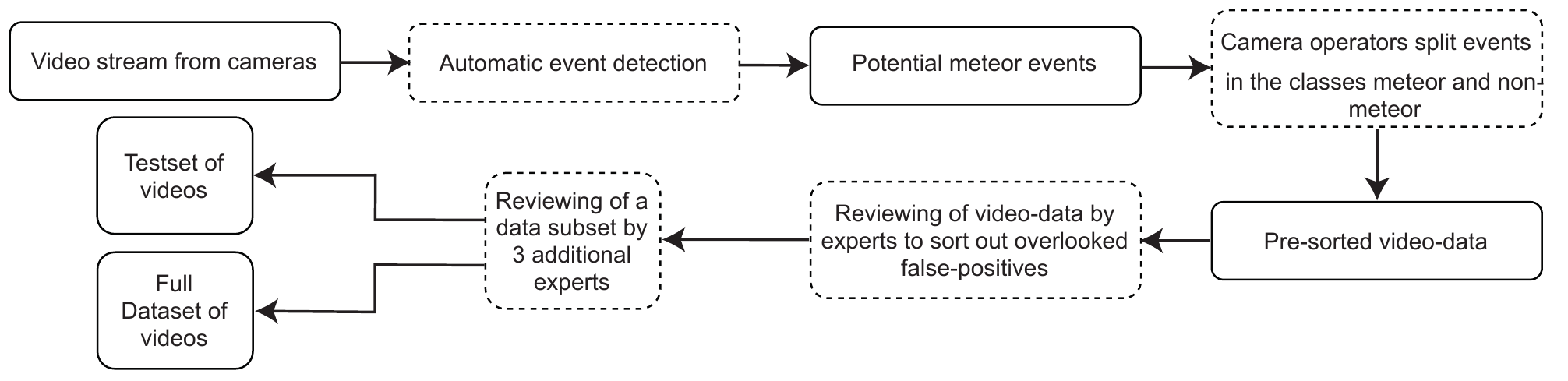}
    \caption{Illustration of Allsky7's data collection process.}
    \label{fig:collecting}
\end{figure*}

\textbf{Process.} Figure~\ref{fig:collecting} depicts Allsky7's data collection process. The automatic event detection\footnote{\hyperlink{https://github.com/
mikehankey/amscams}{https://github.com/
mikehankey/amscams}} detects any movement in the sky based on a simple rule-set that Allsky7 empirically developed to detect every event assiduously. The system then classifies these events and passes them to the camera operator. Afterward, each camera operator reviews these detections to sort out any false-positive events. To ensure a high quality of the dataset, the correct labeling was reviewed again by members of the association Astronomiemuseum-Verein Sonneberg. Samples falsely labeled as positive by the automatic detector are retained for NightSkyUCP's non-meteors class. This way, only non-meteor data that is similar to meteors is included in the non-meteor class. 

Such false-positive labeled samples occur due to local events near the camera stations, such as flying birds, light flares, waving trees, fog or rain, and non-local events far away from the camera stations, such as sky-divers, planes, satellites, flashes of lightning or clouds. While the former is potentially automatically eliminated by geometric trajectory solution constraints derived from footage captured by multiple camera stations, it is impossible to perform this analysis for the latter case since those will be observable across different stations at the same time. Furthermore, the camera systems can not rely on information that depends on shared information in the early stages of the classification process since the events are too numerous, and the synchronization would, therefore, exceed the bandwidth and computation capabilities of the Network.

In this way, we created a dataset  consisting of $10,000$ events in the form of spatially and temporally cropped videos per class \textit{meteor} and \textit{non-meteor}, thus comprising a total of $20,000$ events. A subset of 116 non-meteor events was manually labeled and divided into the subclasses \textit{clouds}, \textit{birds}, \textit{planes}, \textit{trees}, \textit{rain} and \textit{light flashes}. While 116 labeled non-meteor events are not enough to actually train a machine learning algorithm to divide these subclasses, this subset can be used for the evaluation of feature learning and anomaly detection methods of the non-meteor data.

To validate label quality, we drew a random sample of data that we validated against three additional experts' opinions. We report the results of that validation in Section~\ref{sec:limitations}. We further created a test set consisting of 574 samples, 297 being non-meteor and 277 being meteor samples. Five experts classified the samples, and only if all the experts initially decided on the same label, they included the sample into the dataset. \textit{Clouds}, \textit{birds}, \textit{planes}, \textit{trees}, \textit{rain} and \textit{light flashes} often cause false-positive samples; the former three are non-local disturbances and the latter local ones. We labeled 116 non-meteor samples into these six classes to gain insights into the structure of the non-meteor class.

\subsection{Video Preprocessing}
Each event in the dataset consists of the original cropped video, a sum-image, a stack of $32\times32$ pixel crops around the center of gravity of the lightest moving pixels in each frame, and a stack of the corresponding metadata in the form of relative movement and coordinates. Below, we describe the generation of these sum-images and stacks. 

\begin{figure*}
    \centering
    \begin{subfigure}[b]{\textwidth}
        \centering
        \includegraphics[width=\textwidth]{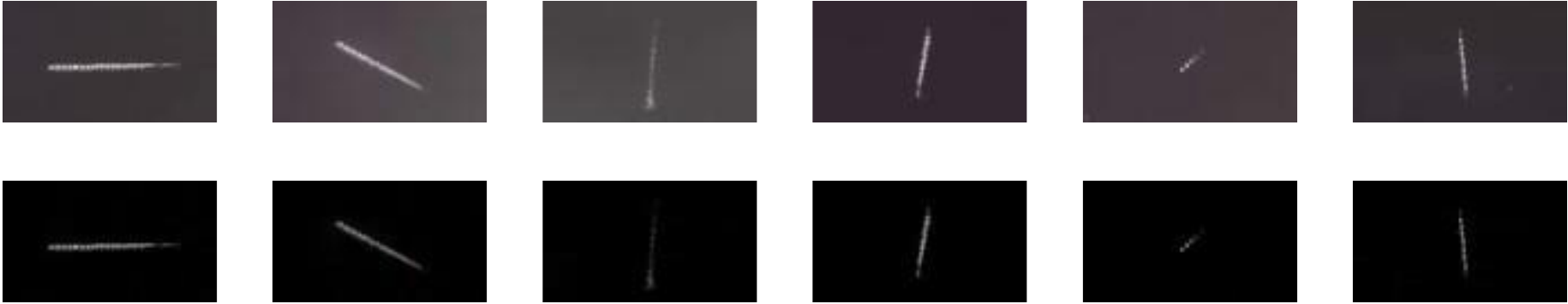}
        \subcaption{Meteors}
    \end{subfigure}
    \begin{subfigure}[b]{\textwidth}
    \vspace{0.5cm}
        \centering
        \includegraphics[width=\textwidth]{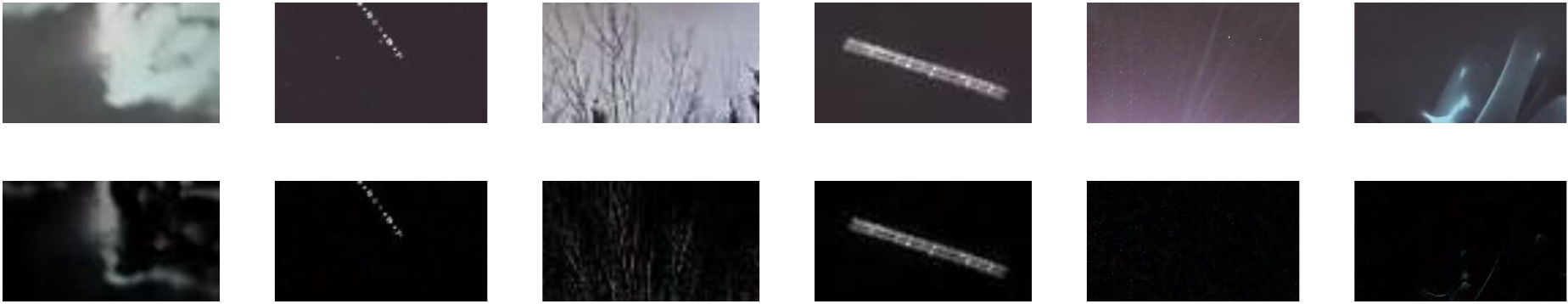}
        \subcaption{Non-meteors}
    \end{subfigure}
    \caption{Examples of sum-images from the two classes meteors (a) and non-meteors (b). In the top rows of (a) and (b) the sum-images before subtracting the first frame are shown and in the bottom row are the sum-images with the first frame subtracted.}
    \label{fig:examples_sumImg}
\end{figure*}

The sum-image per event is generated by comparing each consecutive frame pixel by pixel and by choosing the brighter pixel value, thus taking the maximum temporal pixel value per pixel. This technique is based on \cite{jenniskens2011cams}, but instead of using a fixed number of frames, we use all available video frames to generate our image. Often stars and stationary objects like lanterns that shine brighter than the event disturb the detection of the brightest region. Therefore, we subtract the first frame of the video to remove the non-moving part of the background and to only keep the event's moving pixels yielding reduced noise in the data and improving the detection of the brightest and darkest region in the image. We have made the assumption that the combination of focal length and exposure time of the cameras leads to stars also being perceived as non-moving objects and that the sky's star sidereal motion does not shift more than a pixel over the time of our video. This assumption has been confirmed in practice. Figure~\ref{fig:examples_sumImg} depicts examples of the sum-images with and without subtraction for both classes. In every instance, the upper row is lower in contrast, and bright regions often conceal the region of interest.

\begin{figure}
    \centering
    \begin{subfigure}[b]{0.5\textwidth}
        \centering
        \includegraphics[width=\textwidth]{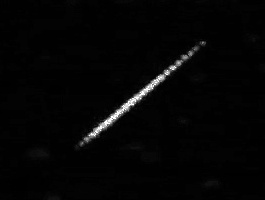}
        \subcaption{Sum-image}
    \end{subfigure}
    \hfill
    \begin{subfigure}[b]{0.45\textwidth}
        \centering
        \includegraphics[trim=12cm 0 0 0, clip, width=\textwidth]{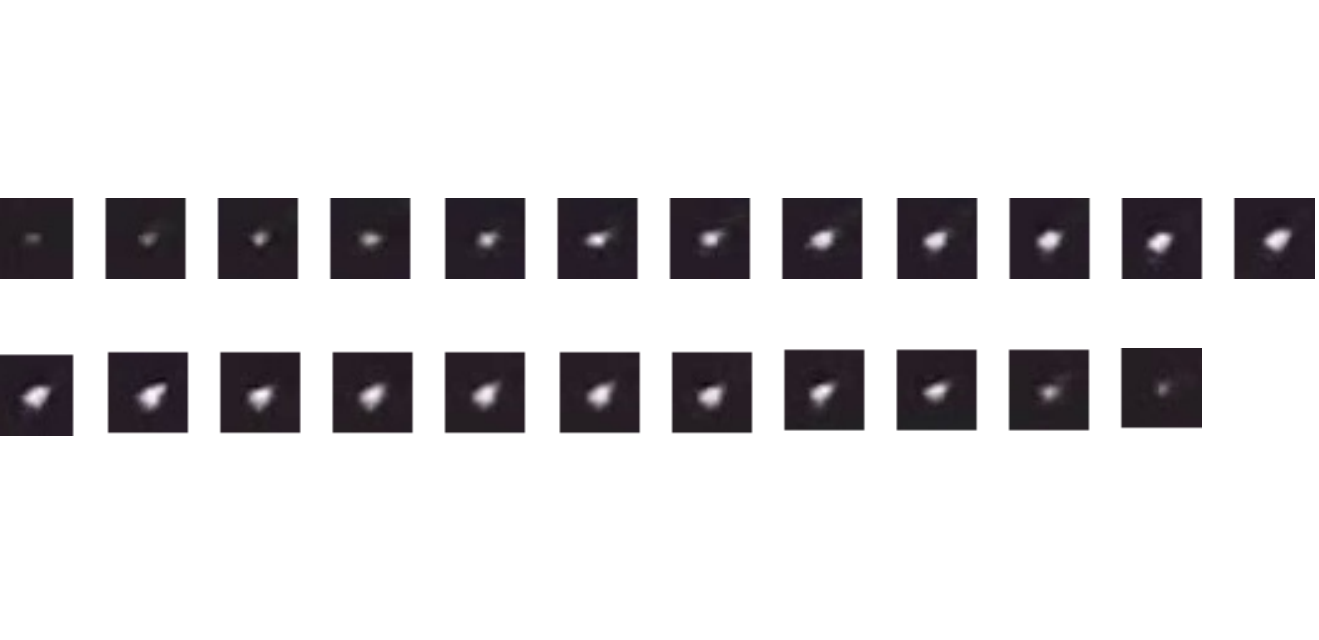}
        \subcaption{Stack of crops}
    \end{subfigure}
    \caption{Example of a generated stack of crops. In (a) the meteor is depicted as sum-image. In (b) the corresponding stack of crops can be seen.}
    \label{fig:imagestack}
\end{figure}

For the extraction of image-stacks, i.e., image-sequence-stacks, we must ensure that a meteor is always the brightest moving pixel cluster in the respective video. The point around which to crop must first be detected to generate the stack of crops for each video frame. We calculate this point based on the brightness of the video frames' pixels. Bright stars or regions infer with the detection based on the brightest area—for example, an incorrectly detected bright star as the center of the region of interest. Therefore we subtract the first frame of the video from each of the following frames before determining the brightest pixel. In this way, we exclude the static background of the video in the search for the meteor. To prevent detecting brightness fluctuations in the image as part of the meteor, we introduce a threshold value (i.e. $10$ of $255$) for the brightness value of the brightest pixel found. If the pixel value is below this threshold, we assume it does not belong to the searched meteor. Because all the cameras are calibrated and dynamically adjust to brightness, this fixed threshold can be used to exclude background noise. We take a $32\times 32$ pixel area crop around the position of the greatest pixel-value above the threshold. We give an example for such a stack of crops in Figure~\ref{fig:imagestack}. For each of the crops, the motion of the brightest pixel between the consecutive frames and the coordinates relative to the video is stored.

\subsection{Details of the Dataset}
\begin{figure}
	\centering
	\begin{subfigure}[b]{0.49\textwidth}
		\centering
		\includegraphics[width=\textwidth]{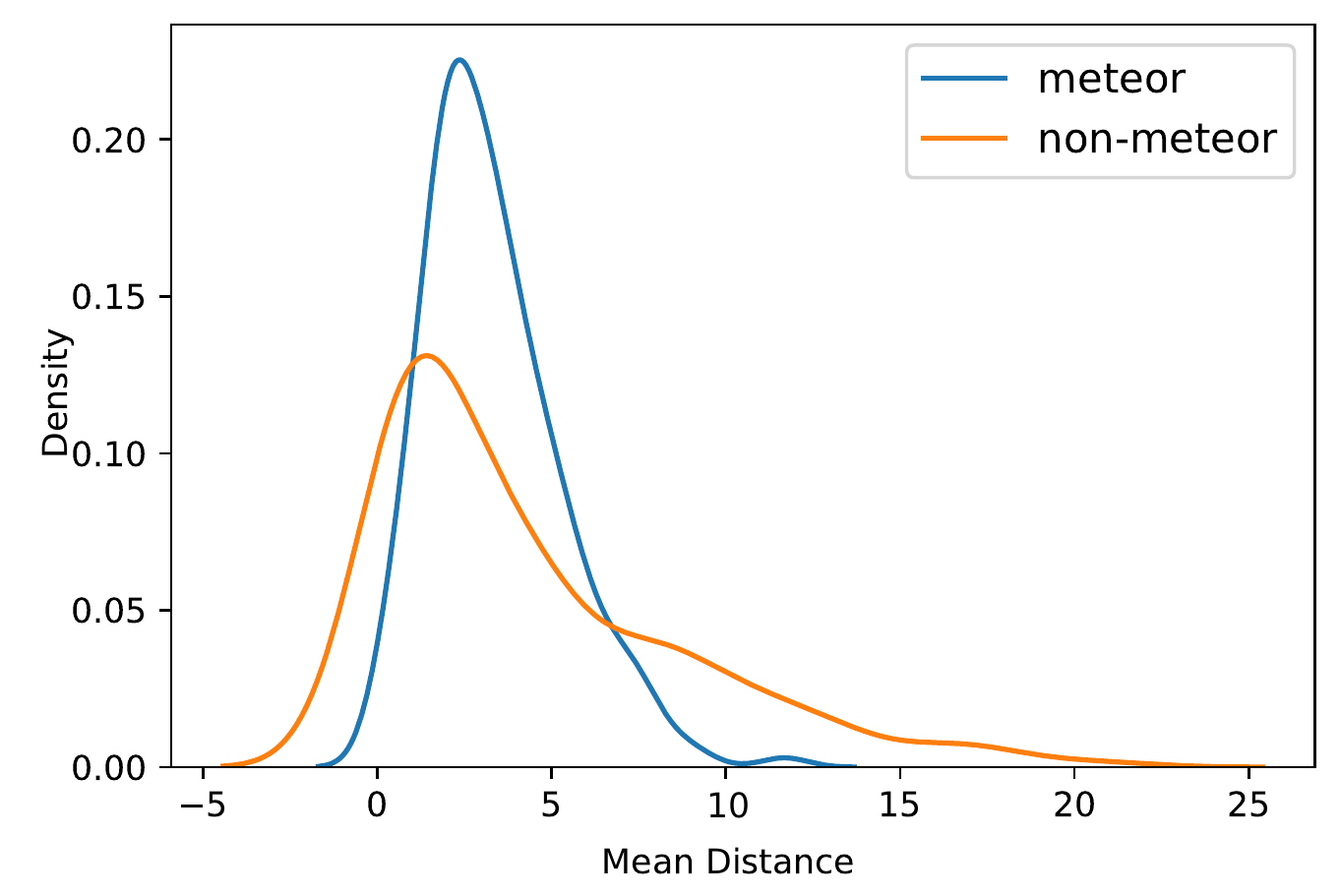}
		\caption{Euclidean distance}
	\end{subfigure}
	\hfill
	\begin{subfigure}[b]{0.49\textwidth}
		\centering
		\includegraphics[width=\textwidth]{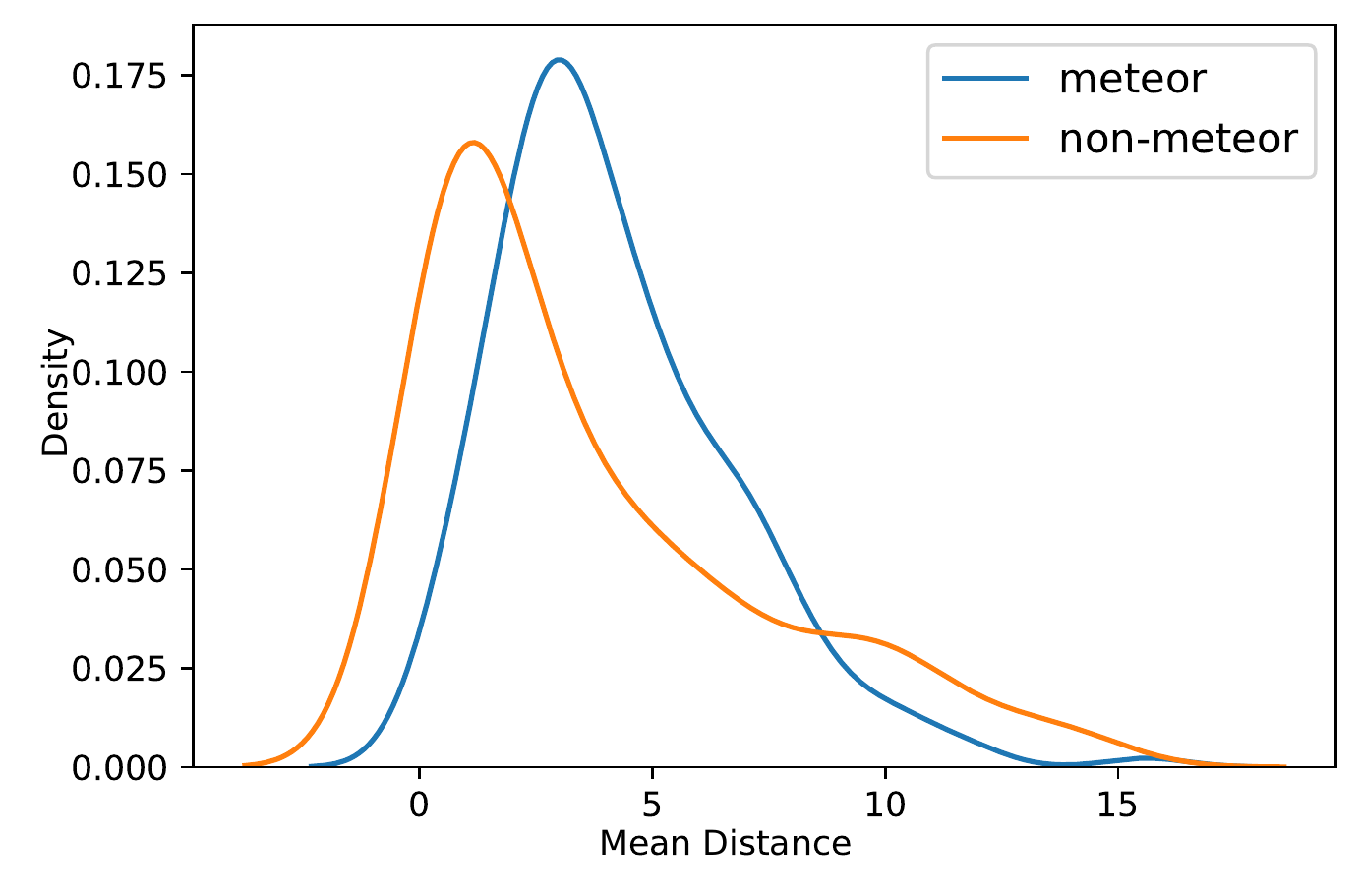}
		\caption{Manhattan Distance}
	\end{subfigure}
	\caption{Display of the average distance traveled per frame for all events in the dataset, divided by class. The visualization is done by a kernel density estimation (KDE). The Euclidean distance (a) and the Manhattan distance (b) serve as distance measures.}
	\label{fig:hist_dist}
\end{figure}

The size of the videos varies between $80\times44$ pixel and $640\times360$ pixel. The length of the videos varies between 0.08 and 14.32 seconds. Respectively, the number of frames for each event is different, ranging from 2 to 358 frames. That means some events are only visible in a blink of an eye while others last comparably long. The longest and brightest events, so-called fireballs, are the most exciting events in the sky. To understand the structure of the data better, we performed an analysis. We were interested in the average speed of the events and the average pixel distance since it is reasonable to assume a different pixel distance traveled per frame for meteors and non-meteors. Figure~\ref{fig:hist_dist} shows the distribution of Euclidean and Manhattan distances events traveled per frame over the whole dataset estimated by a kernel density estimation. One can observe that the non-meteors distribution has a broader range with a shorter mean distance and a long tail towards long distances. 

\begin{figure}
	\centering
	\includegraphics[width=0.5\textwidth]{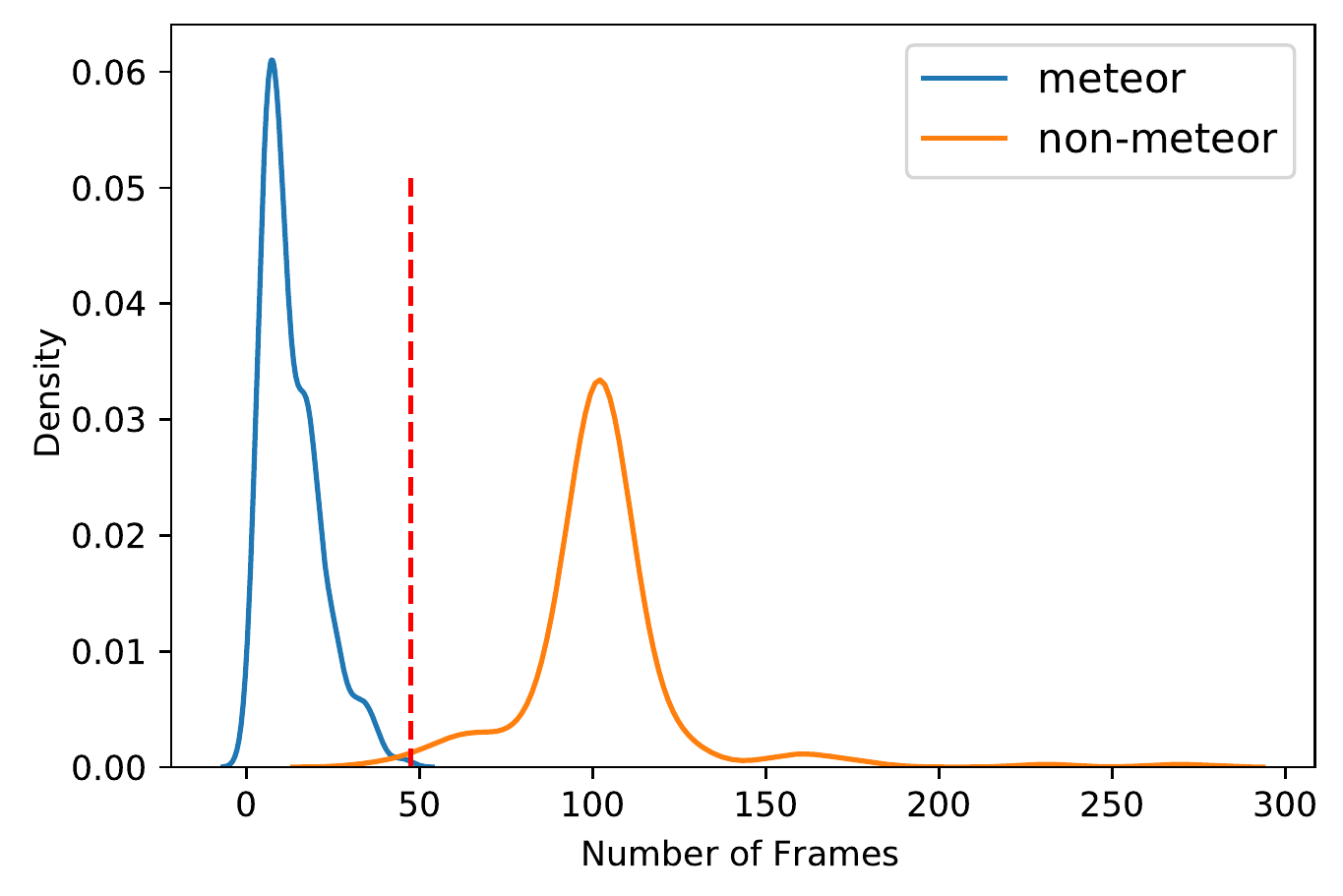}
	\caption{Plot of the average event length per class. The visualization is done with a kernel density estimation (KDE). In red is the optimal threshold for separating the two classes based on the number of frames.}
	\label{fig:event_len}
\end{figure}

In the same way, Figure~\ref{fig:event_len} shows the distribution of the event's duration estimated by a kernel density estimation. Here we counted the number of frames in which the brightest pixel is above a threshold of $10$. The figure illustrates that non-meteor events, with $100$ frames average, are much longer than meteor events, whose average length is $13$ frames. Since the two classes slightly overlap, we calculated a threshold that separates them from the distributions. We obtained the best separation with a threshold of $47.5$. Applying this threshold, we classify $395$ samples wrong, leading to an accuracy of $98.02\%$. This accuracy could act as a baseline for classification performance.
We assume that the difficult and interesting events are mainly those which lie at the intersection. Especially very bright meteors, so-called bolides or fireballs, usually last much longer than the average meteor event and would be assigned to the class of non-meteors by this simple classification rule.


%% file: content/applications.tex

%% file: content/experiments.tex

\section{Experimentation}\label{sec:experiments}
These experiments provide insight into the data's structure, also representing proof of concept for the different tasks we described before. Although we compared different methods and evaluated a specific range of parameters, many choices were made by best practice and may not be optimal in the specific setting. We, however, provide the implementation of four experiments to aid scientists performing Wide-area-sky-monitoring and give baseline results to compare with:
\begin{enumerate}
    
    \item \textbf{Classification.} Classification refers to the task of predicting a class label for a given input sample. NightSkyUCP allows classifying meteors and non-meteors based on the sum-images, the videos directly or from the crop- and the motion-stacks. An additional problem is performing the classification exclusively based on the first few, e.g., five, frames of an event. The advantage of classifying only on the first frames would be that waiting for the whole meteor in a live video stream would not be necessary. This way, live classification could be performed with only a little delay.
    \item \textbf{Feature learning.} NightSkyUCP samples can be used to learn features that allow for clustering of the non-meteor data. It would be interesting to find whether the different types of non-meteors can be clustered in an unsupervised manner. A small subset of non-meteor data is already labeled to check the quality of such clusters.
    \item \textbf{Anomaly detection.}  Researchers could also perform anomaly detection on the meteor data. The task there should be distinguishing between meteor and non-meteor events while only training with meteor data. Especially, non-meteors like airplane flash chains look very similar to meteors. It is to be investigated whether the positive class can be defined precisely enough by the data of the meteor class to detect the more difficult non-meteors correctly. 
    \item \textbf{Extrapolation.} The motion-stacks can be uses to predict the path of a meteor.  Nevertheless, it is unclear how curvature and motion are connected to the brightness in each frame. Machine learning could find a connection between the brightness and the motion so that researchers get insights into the type of an event. Often, meteorites are covered by clouds or trees in the foreground so that their path is not complete and their trajectory is unclear. Meteor data could be artificially covered to learn how to recover the lost information regarding the path and brightness.
    
\end{enumerate}

\subsection*{Experiment i: Classification}
\label{sec:class}
The first task is classifying data into two classes, meteor and non-meteor, based on the sum images. We trained an ImageNet pre-trained ResNet20 \cite{He2016} in combination with a spatial pyramid pooling (SPP) layer \cite{He2015} allowing for processing different input sizes in a convolutional neural network with a linear classifier head. We employed a pre-trained model because it is already trained on an extensive dataset to generate well separable features. Since it is usually a complicated procedure to tune all network architecture parameters to fit the data well enough, we leave it to future work to find such an architecture for meteor data. We used a batch size of $100$, the Adam optimizer \cite{Adam2014} with a learning rate of $0.005$ and trained for 200 epochs. We also use synaptic scaling loss as regularization technique \cite{Hofmann2021} with a scaling rate $\gamma=1e-7$ and a target activation of $z_T=0.5$ together with batch normalization as part of ResNet20 \cite{BatchNorm2015} and exponential dampening of the weight updates using $\mu=0.9999$. We split the data randomly into train and test data, using $80\%$ for training and $20\%$ for evaluation. For robust evaluation, we employ a Monte Carlo cross-validation with five iterations. Across these five training iterations, we achieved a mean accuracy of $97.33\%$ with a standard deviation of $0.6\%$. This result is not considerably below the baseline of $98,02\%$ that we achieved by classifying based on the event length. We, therefore, conclude that the sum-images without temporal information are not sufficient for reliable classification.

To check if the explicit temporal information, motion, and direction contained in the image- and motion-stacks and the metadata are better suited for the task of classification, we conducted further experiments with different feature extraction methods.
These methods get both the stack of metadata and the imagestacks as input, thus containing more temporal infomation than the sum-images.
As methods for the feature extraction, we compared an LSTM (Long short-term memory), GRU (gated recurrent unit), TCN (Temporal Convolutional Network), TDNN (Time Delay Neural Network) and an Autoencoder based on LSTMs. 
Furthermore, we compared different classifiers, specifically a fully connected layer, SVM (Support Vector Machine) and Random Forest.
The hyperparameters used for the different methods can be found in appendix B.
Again we split the data randomly into train and test data using $80\%$ for training and $20\%$ for evaluation and employ a Monte Carlo cross-validation with five iterations.
The experimental results can be seen in table \ref{tab:resultsEval}.
We used the most recent version of pytorch for our experiments. On a NVidia 2080Ti, depending on the used method, training a network for classifying took between 22 minutes and 5 hours

\begin{table*}
    \centering
    \small
    \begin{tabular}{rcccccc}
    \hline
         & \textbf{LSTM} &  \textbf{GRU} &   \textbf{TCN} &   \textbf{TDNN} &   \textbf{SPP} &   \textbf{AEC}  \\ \hline
         & ~~~Acc$\pm$STD &   ~~~Acc$\pm$STD &  ~~~Acc$\pm$STD &  ~~~Acc$\pm$STD & ~~~Acc$\pm$STD &  ~~~Acc$\pm$STD   \\ \hline
        Linear       & 99.00 $\pm$ 0.13 & 99.05 $\pm$ 0.08 & 98.89 $\pm$ 0.15 & 94.38 $\pm$ 0.24 & 97.52 $\pm$ 0.24 & 99.01 $\pm$ 0.15 \\ \hline
        SVM            & 99.03 $\pm$ 0.10 & 99.10 $\pm$ 0.09 & 98.88 $\pm$ 0.15 & 94.62 $\pm$ 0.21 & 97.83 $\pm$ 0.16 & 98.99 $\pm$ 0.09 \\ \hline
        RF  & 99.03 $\pm$ 0.14 & 99.09 $\pm$ 0.09 & 98.99 $\pm$ 0.18 & 95.96 $\pm$ 0.18 & 97.64 $\pm$ 0.13 & 98.92 $\pm$ 0.09 \\ \hline
    \end{tabular}
	\caption{Accuracy and standard deviation in percent for the evaluation of feature extractor and classifier combinations. Shown are the mean and standard deviation of the accuracy over 10 training runs on random data splits.}
	\label{tab:resultsEval}
\end{table*}

The best mean accuracy of $99.1\%$ can be achieved with GRU as feature extractor and SVM as classifier. The best overall performance is achieved with GRU as feature extractor and SVM as classifier. Although the SVM classifier achieves an equivalent accuracy on LSTM and GRU, we consider GRU superior due to a reduced number of learnable parameters and a faster inference time. The achieved accuracy is better than the baseline of $98.02\%$ using a rule based on the length of the events. Future iterations of this dataset provided at the given link will contain a greater number of those samples. Our classifier achieved an average accuracy of $83.54\%$ on those events. It is noticeable that this accuracy is far below the mean value of the accuracy overall test data. This decrease in accuracy confirms the assumption that these samples are challenging to classify. Furthermore, we assume that the length of the event is also included in the features extracted by GRU. We also considered the data that the trivial classification rule would incorrectly classify for each run for the combination of GRU and SVM.

\begin{figure}
    \centering
    \includegraphics[trim=0 0 5 0, clip,width=1.00\linewidth]{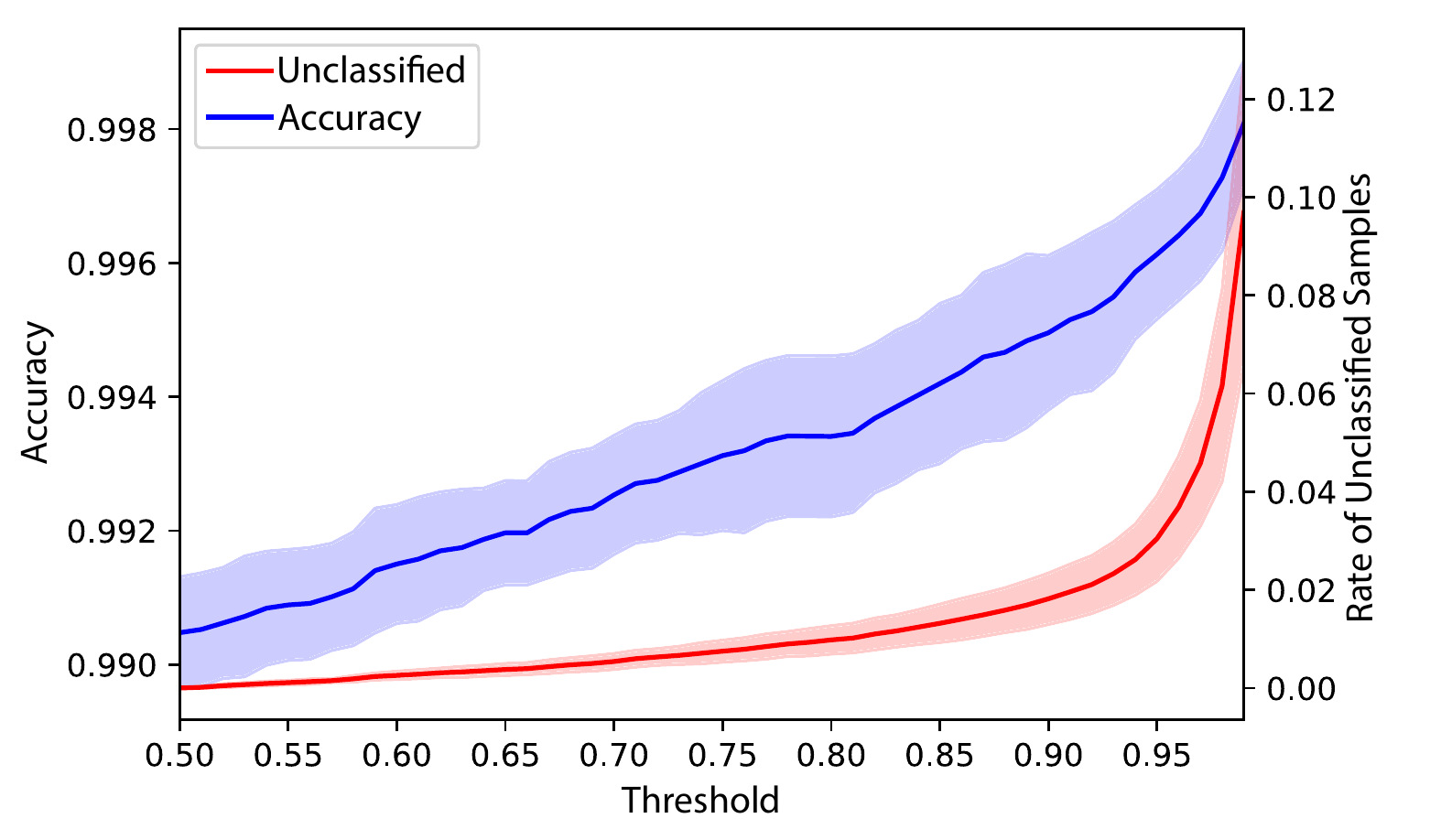}
    \caption{This figure presents the accuracy and the corresponding rate of unclassified samples when tuning the classifier threshold.}
    \label{fig:acc_fnr}
\end{figure}

Figure \ref{fig:acc_fnr} presents the accuracy and the corresponding rate of unclassified samples when tuning the classifier threshold - a neurons' minimum activation that it must exceed. Increasing the threshold improves the accuracy and increases the estimated proportion of samples that have to be reviewed by specialists. Nevertheless, samples with lower scores represent more interesting samples with a higher value for training 
\cite{visapp19}. We, therefore, are gathering more and more informative samples and increasing the accuracy to 99.98 percent while decreasing the workload by the factor of 8-10.

\subsection*{Experiment ii: Feature Learning}
\begin{figure*}[h]
	\centering
	\begin{subfigure}[b]{0.47\textwidth}
	
		\centering
		\includegraphics[trim={15 10 20 50},clip,width=\textwidth]{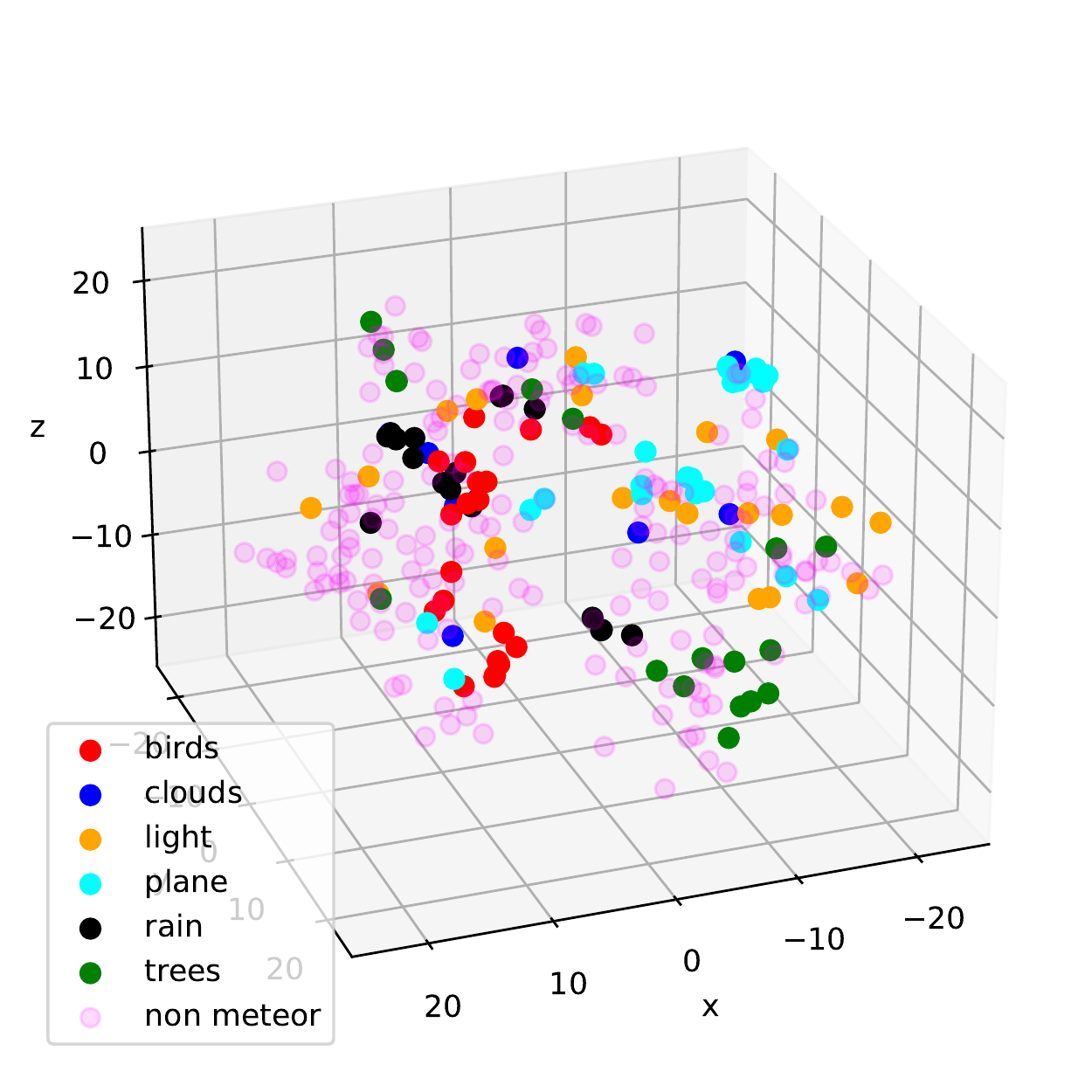}
		\caption{ Data manifold visualization using three components \textit{t-SNE} with highlighted sub-classes.}
		\label{fig:cluster_tsne}
	\end{subfigure}
	\hfill
	\begin{subfigure}[b]{0.47\textwidth}

		\centering
		\includegraphics[trim={20 20 15 5},clip,width=\textwidth]{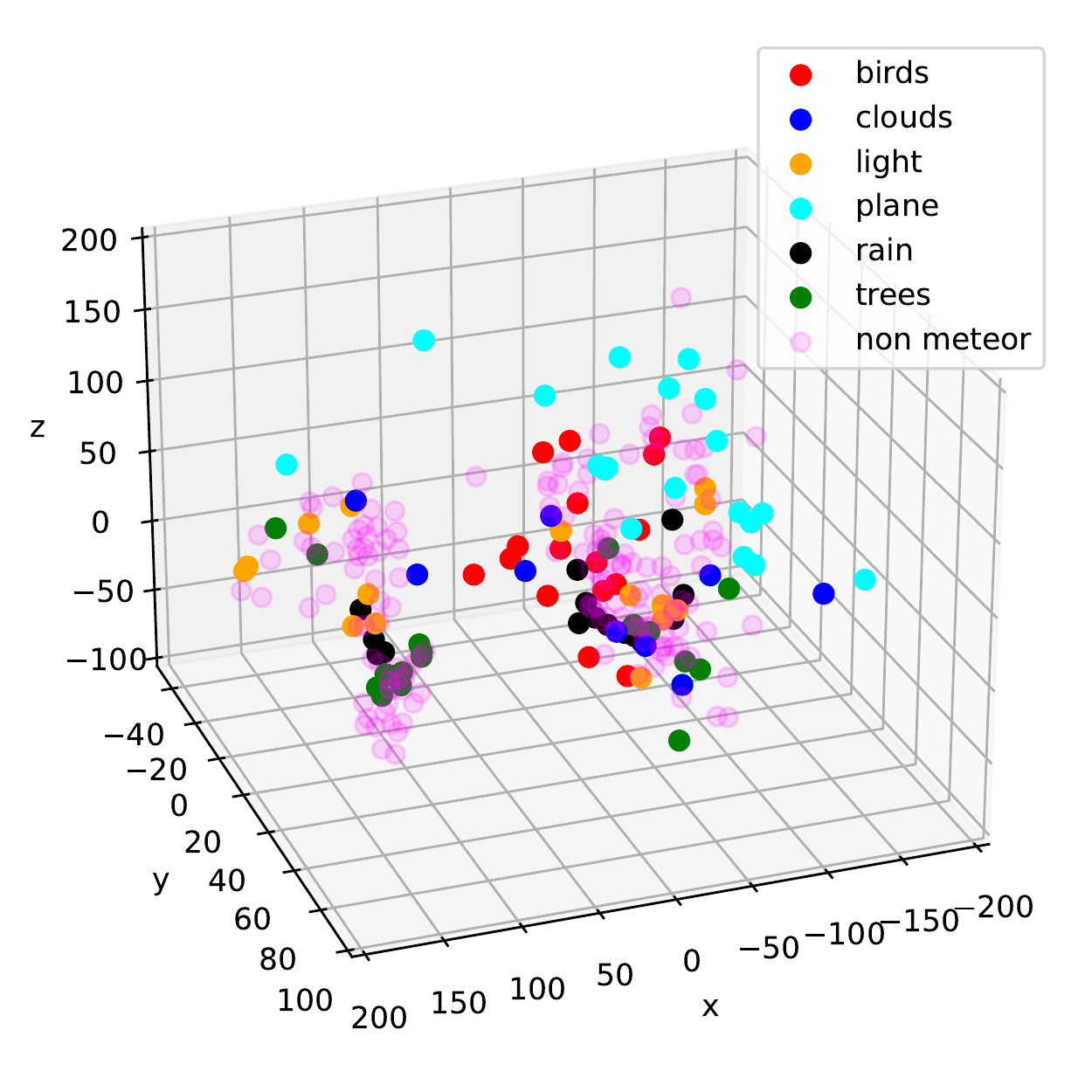}
		\caption{Three components Isomap trained on twenty percent of non meteor samples. Highlighted are all samples with sub-classes.}
		\label{fig:cluster_isomap}
	\end{subfigure}
	
	\caption{The scatter plots show non-meteor samples embedded into three dimensions. The features are generated using a trained ResNet20 with SPP-Layer. The labeled sub-classes are highlighted. The colors depict the class non-meteor and the sub-classes \textit{clouds}, \textit{birds}, \textit{planes}, \textit{trees}, \textit{rain} and \textit{light flashes}.}
	\label{fig:cluster}
\end{figure*}

The second task is evaluating how well the non-meteors can be clustered without using the label information during training.
We use the convolutional network trained for Task~1 as a feature extractor for the sum-images, which results in feature vectors of size $10$.
The dimension reduction techniques \textit{Isomap} and \textit{t-SNE} \cite{tenenbaum2000global,VanderMaaten2008} are used to visualize 3D embedding (cp. Figure~\ref{fig:cluster}). While the \textit{Isomap} embedding is fit on a 10\% split of the non-meteor data to show a result in a training/validation setting, the \textit{t-SNE} embedding is fit on all data to visualize the manifold.

Figure~\ref{fig:cluster_tsne} shows that the manifold is split into two seemingly separable clusters, one containing particularly birds and rain and one containing especially trees, lights, and planes. Clouds seem to be equally distributed. Notably, birds are the only class that is completely confined to one cluster. Trees, however, tend to be included in both clusters but seem to form sub-clusters that are a part of the other samples. We can also see that planes and lights seem to be sparsely distributed over one cluster, hardly separate from each other. The 3D \textit{Isomap} embeddings (cp. Fig.~\ref{fig:cluster_isomap}) trained only on 10\% of the non-meteor data shows the same two clearly separable clusters as the \textit{t-SNE}. Here, the samples belonging to the tree class seem to form one cluster while birds, clouds, and planes form a second cluster with nearly no sample belonging to the first cluster. Furthermore, plane embeddings seem separable from trees and clouds but highly overlap with birds and lights. Here, results of \textit{t-SNE} and \textit{Isomap} tend to be contradicting. While planes and lights seem to be nicely separable using \textit{Isomap}, \textit{t-SNE} shows that they are mixed up.

We trained the network used for the feature extraction to separate the meteor and non-meteor classes, so it is not surprising that the learned features are insufficient to separate the non-meteor sub-classes. However, this experiment shows that some kind of separation is possible even with no specialized features, allowing for some insights into the data. For future work, there is considerable potential to achieve good clustering results. It is needed to train networks like siamese networks for further insights into the inner data fabric \cite{triplet_loss_2010} \cite{hoffer2015deep}, \cite{chen2020simple}. 

\subsection*{Experiment iii: Anomaly detection}
The anomaly detection experiment is conducted using the auto-encoder setting (cp. \ref{sec:class}) trained for 200 epochs on a subset of 9000 meteor samples. The remaining 1000 samples are divided into two parts of 500 samples each. The first part is used for training a k-nearest neighbors (k-NN) model, and the second for its evaluation. We evaluated the average Euclidean Distance of each sample to its five nearest neighbors. The anomaly detection is performed on the remaining 10000 \textit{Non\_Meteor} samples. Averaged over all samples, the distance for the evaluation split was 0.53$\pm$0.41 that is significantly different from the distance of 1.47$\pm$0.52 observed for the \textit{Non\_Meteor} split. 

\begin{figure*}
	\centering
	\begin{subfigure}[b]{0.45\textwidth}

		\includegraphics[trim={0 0 0 0},clip,width=\textwidth]{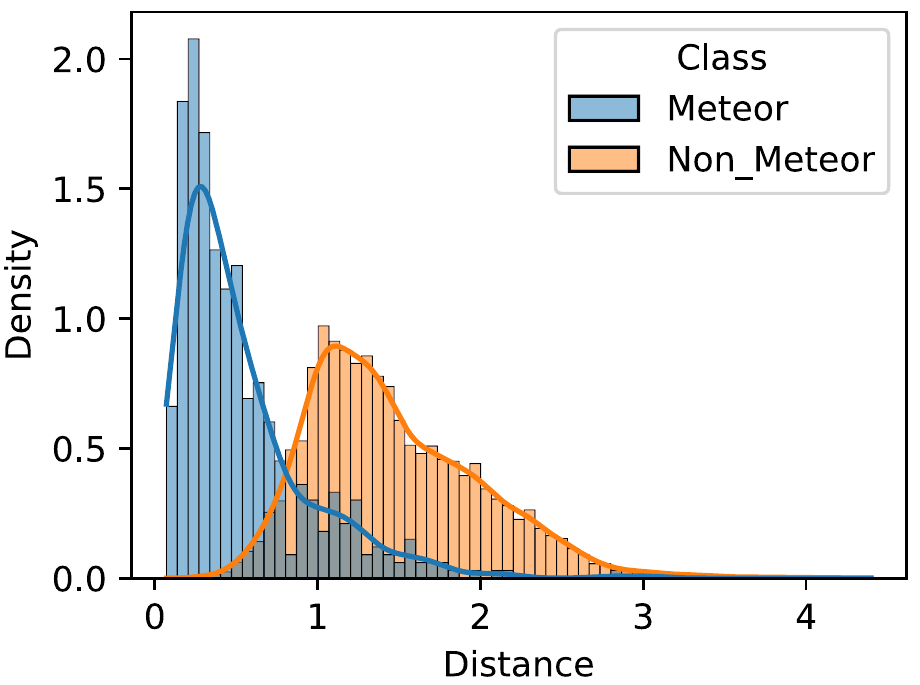}
		\caption{Kernel Density Estimation and Histogram of the k-NN distances for the classes \textit{Meteor} and  \textit{Non\_Meteor}.}
		\label{fig:anomaly_dist}
	\end{subfigure}
	\hfill
	\begin{subfigure}[b]{0.48\textwidth}
		\centering
		\includegraphics[trim={0 0 0 0},clip,width=.7\textwidth]{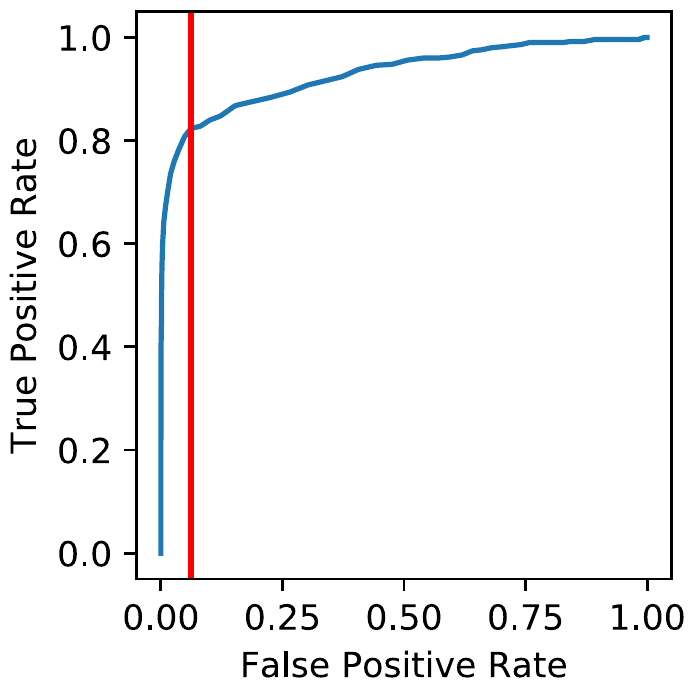}
		\caption{Receiver Operating Characteristic (ROC) curve of the k-NN anomaly detection. The vertical red line denotes a distance of 0.85. TPR and FPR are 82 \% and 6\% respectively.}
		\label{fig:anomaly_roc}
	\end{subfigure}
	
	\caption{Visualization of the classes' distributions and the Receiver Operating Characteristic (ROC) for experiment on anomaly detection.}
	\label{fig:cluster_anomaly}
\end{figure*}

Figure \ref{fig:anomaly_dist} illustrates that the distribution of the \textit{Meteor} and \textit{Non\_Meteor} classes overlap slightly and both have their center of gravity about 2 standard deviations distant from each other. We, therefore, utilize a threshold to separate the classes. As figure \ref{fig:anomaly_roc} shows, a threshold of 0.85 gives the best results.

Although the classes have significantly different embeddings, they overlap too much, and the observed False Positive Rate of 6\% is too large to be operational. We, therefore, conclude that more refined methods are required to solve the problem of missing \textit{Non\_Meteor} samples if none were available. Nevertheless, anomaly detection methods trained with our model can still provide a ranked candidate list for outliers and anomalies that do not fit the meteor data distribution, like non-meteors we already captured in our subclasses or new events like camera artifacts. Moreover, our dataset makes such samples available to train classifiers and evaluate anomaly detection methods.

\subsection*{Experiment iv: Extrapolation}

\begin{figure*}
	\centering
	\begin{subfigure}[b]{0.45\textwidth}
		\includegraphics[trim={50 45 40 40},clip,width=\textwidth]{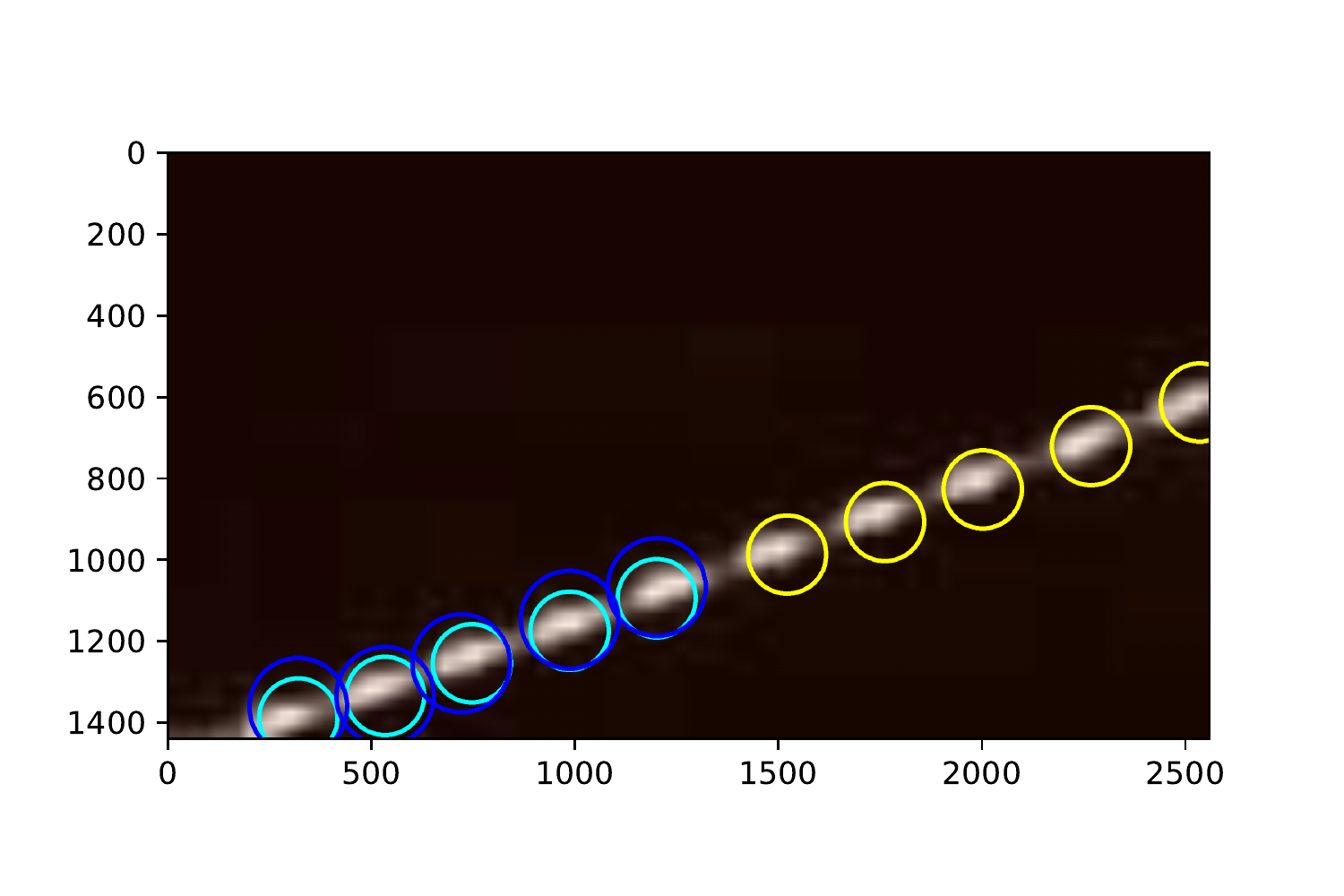}
		\caption{Extrapolating on a \textit{Meteor} class examples. Five time steps shown. Mean error 0.9 pixels.}
		\label{fig:reg_meteor}
	\end{subfigure}
	\begin{subfigure}[b]{0.45\textwidth}
		\includegraphics[trim={50 45 40 40},clip,width=\textwidth]{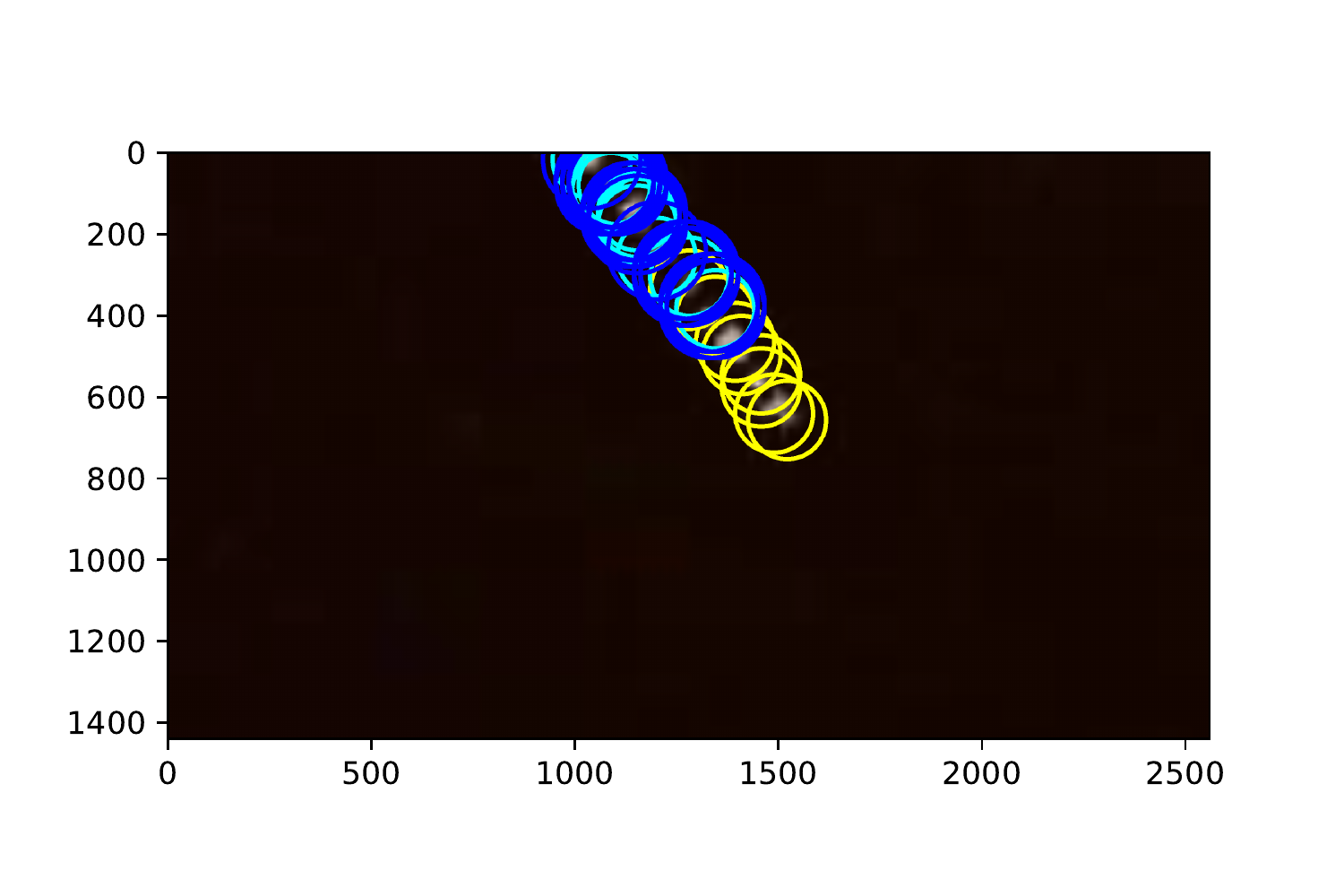}
		\caption{Extrapolating on a \textit{Plane} class examples. 34 time steps shown. Mean error 0.2 pixels.}
		\label{fig:reg_plane}
	\end{subfigure}
	\begin{subfigure}[b]{0.45\textwidth}
		\includegraphics[trim={50 45 40 40},clip,width=\textwidth]{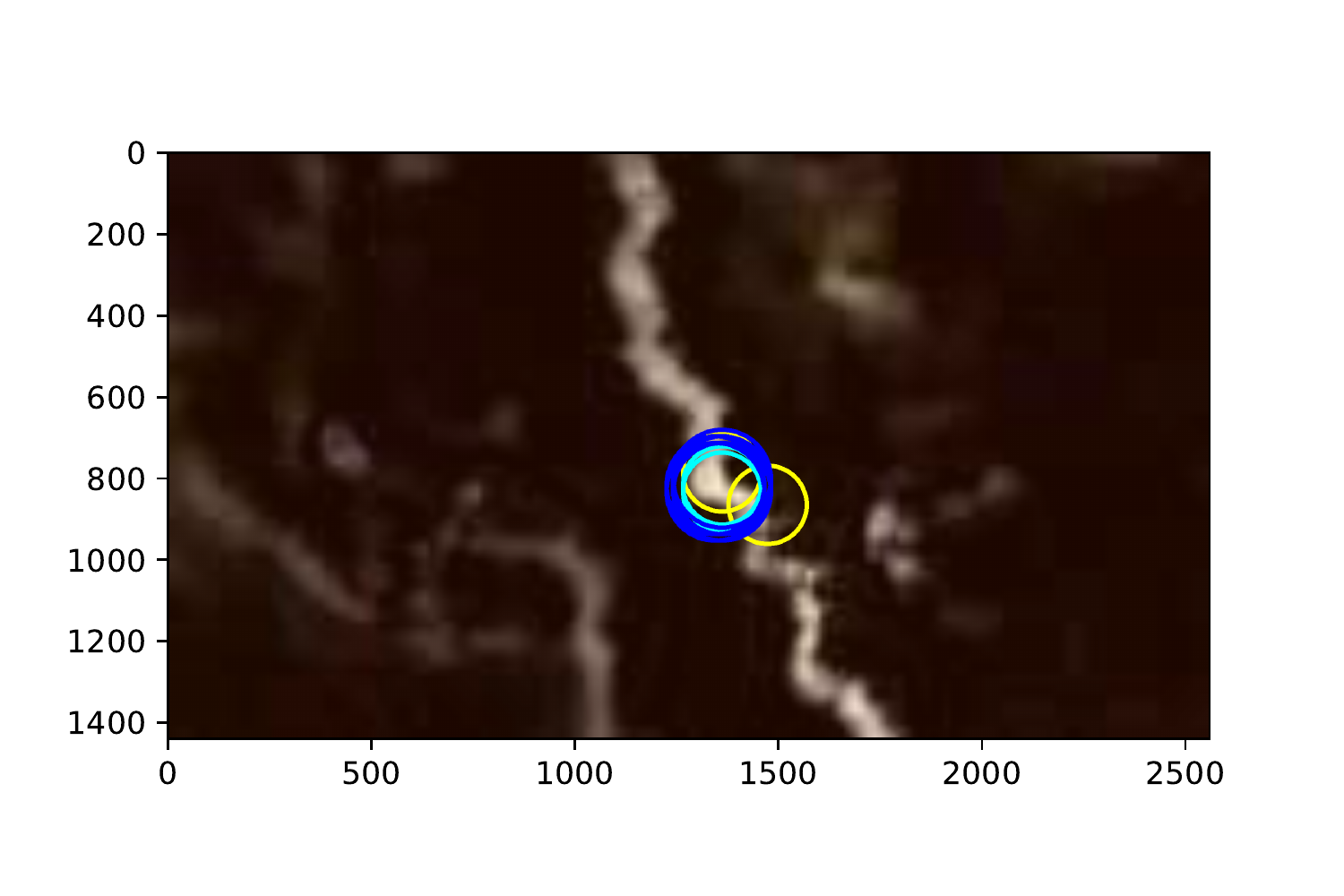}
		\caption{Extrapolating on a \textit{Cloud} class examples. 50 time steps shown. Mean error 1.4 pixels.}
		\label{fig:reg_cloud}
	\end{subfigure}
	\begin{subfigure}[b]{0.45\textwidth}
		\includegraphics[trim={50 45 40 40},clip,width=\textwidth]{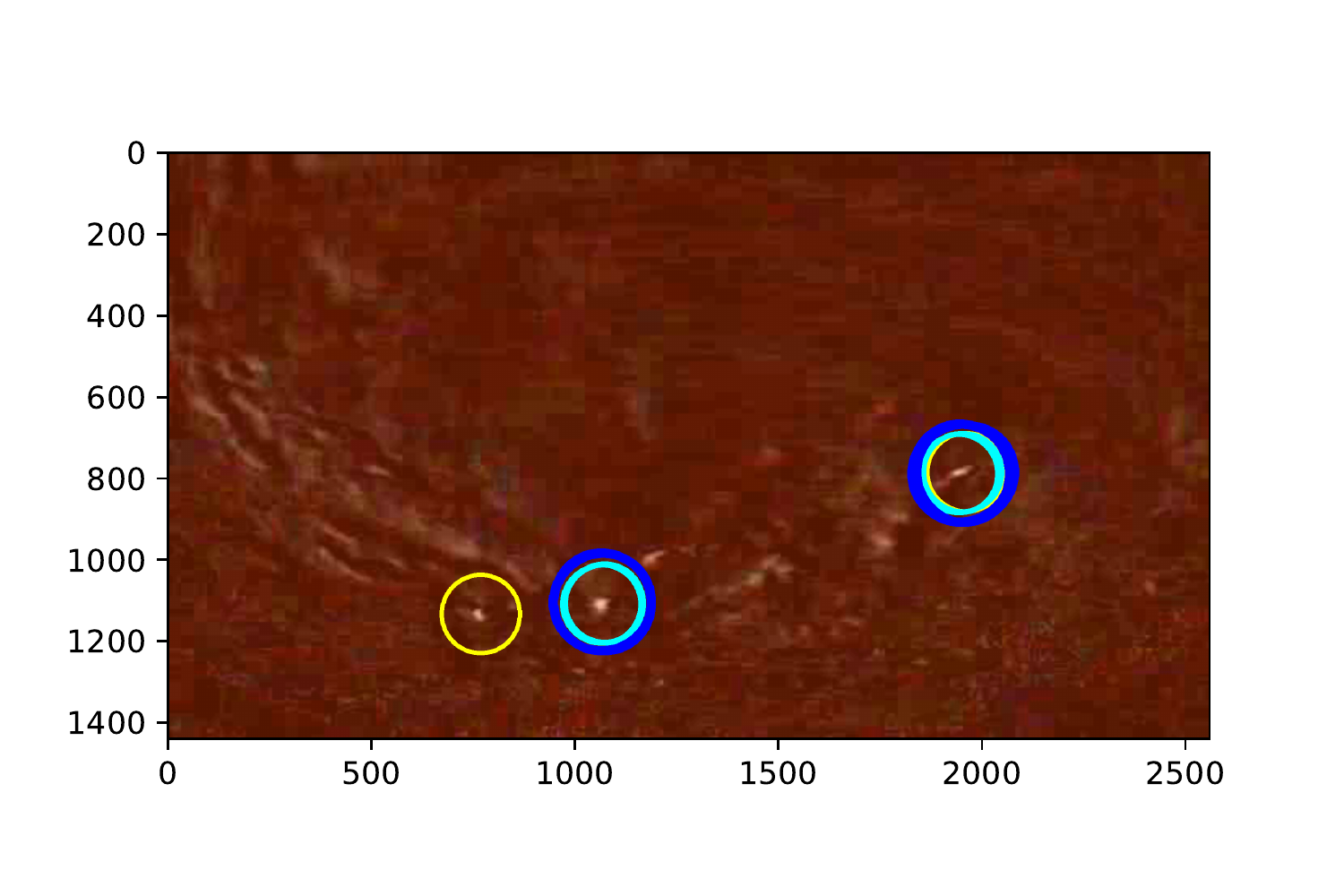}
		\caption{Extrapolating on a \textit{Bird} class examples. 74 time steps shown. Mean error 0.3 pixels.}
		\label{fig:reg_bird}
	\end{subfigure}
	\begin{subfigure}[b]{0.45\textwidth}
		\includegraphics[trim={50 45 40 40},clip,width=\textwidth]{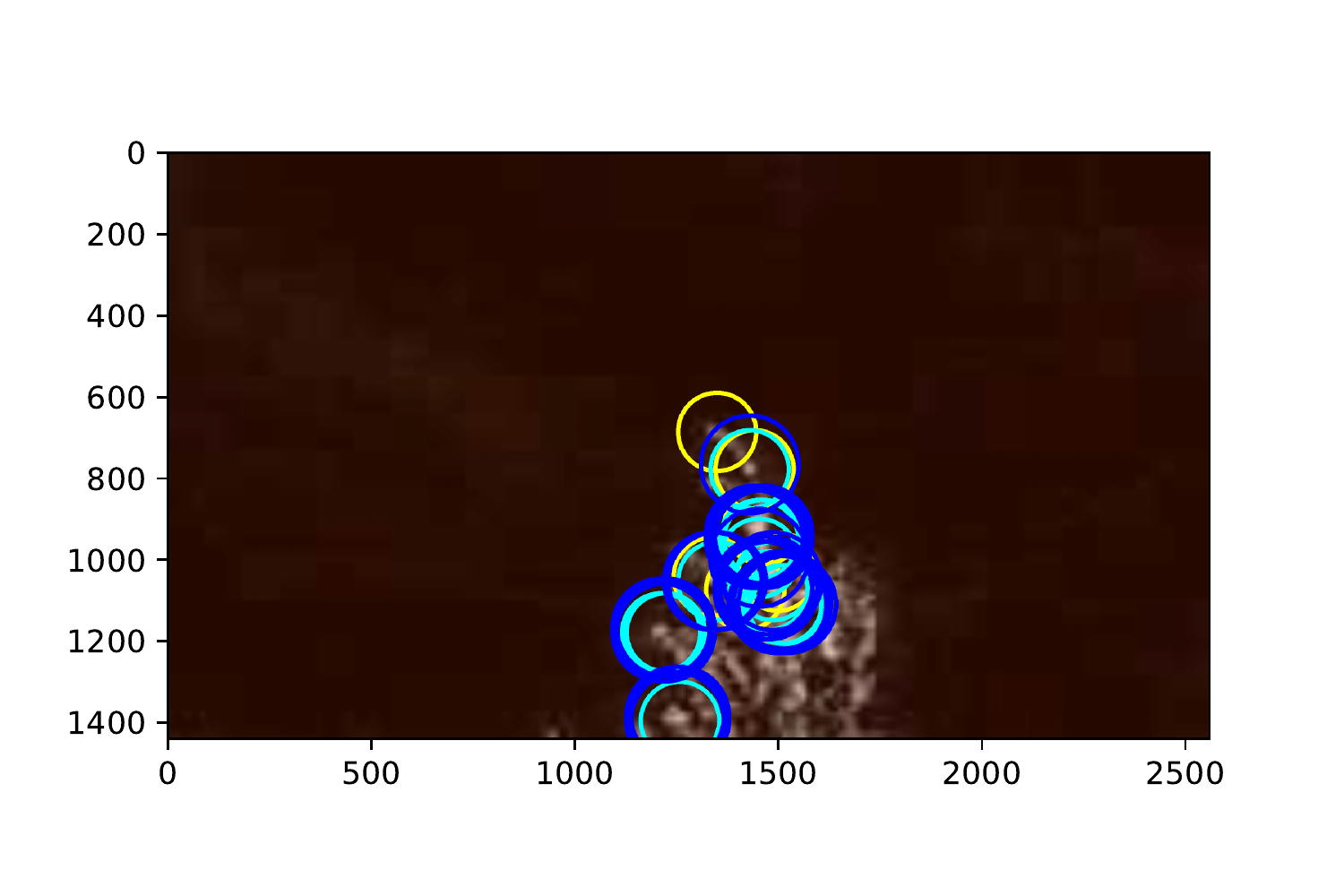}
		\caption{Extrapolating on a \textit{Tree} class examples. 52 time steps shown. Mean error 0.17 pixels.}
		\label{fig:reg_tree}
	\end{subfigure}
	\caption{Examples of extrapolations from the first halves of detected sequences chosen from the test set. The yellow color depicts the input sequence pixel coordinates, the light blue depicts the ground truth and dark blue the predicted pixel coordinates. Halve the sequence length was input and halve was extrapolated.}
	\label{fig:reg}
\end{figure*}

We experimented on extrapolation using a simple transformer network \cite{transformer_2017}.  The input samples consist of two coordinates of the brightest pixel per frame, together with the respective velocities. The transformer's output is fed into another linear layer with four output neurons. Since the input dimension of four was too small for the transformer network to learn, we decided to increase the dimensionality with one linear layer with 64 neurons. We trained the model using an  L2 regression loss. Since we did not use any regularization like gradient clipping, the transformer started with a high error of 2716236 square pixels and needed about 1000 epochs to get to an error lower than ten. After 1200 epochs, the training ended at an error rate of 0.73 square pixels. The training needed 22 days of computation time on a NVIDIA3080Ti GPU. Figure \ref{fig:reg} exemplarily shows the sum-images of samples we used in the evaluation. All pieces belonged to the test set and were not seen during training. Most samples in the test set belong to the classes meteor and plane. Figures \ref{fig:reg_meteor} and \ref{fig:reg_plane} show their path through the field of view. One can see the object paths' slight curvatures that the model precisely predicted.
Moreover, the accuracy diminishes if the object is faster or changes its velocity. Therefore the meteor class shows a larger error than the plane. Figures \ref{fig:reg_cloud}, \ref{fig:reg_bird} and \ref{fig:reg_tree} show slow or stationary objects. Here the bird and the tree show the minor error due to the considerable sequence length and the mostly static brightest pixels detected in the camera view. The most significant error is observed for the cloud sample with noisy detections with no clear direction.

%% file: content/limitations.tex

\section{Limitations}\label{sec:limitations}
We are aware that data are biased. Therefore, we outline our countermeasures against several types of bias.

\textbf{Selection bias.} 
The NightSkyUCP dataset consists primarily of ordinary meteors and an unbalanced number of subclasses of non-meteors. The dataset includes valid image frames from meteors and non-meteors, therefore non-meteor classes like lens flares or compression artefacts where at this point too few to include as specific sub-class into the dataset. As the classification accuracy improves, such sub-classes will get more prominent and, therefore, included into the dataset's future versions. The unbalanced character of the non-meteor class makes a classification of sub-classes of non-meteors a challenging problem. We collected 20,000 samples from 2019 to 2021 from several stations that covered several meteor showers, different weather situations and seasons. We collected nearly all false positives. However, the dataset lacks samples excluded by our autonomous algorithm in the first place. Because of this, it might not be suitable for the classification of fireballs that are rarely present in the dataset. 

\textbf{Overfitting and underfitting.} Overfitting occurs when the data is easily connected to returning noise based on the collection process. We intend to counter that by collecting samples from different stations that work under different conditions. To prevent underfitting, we collected as much data as possible that means 10,000 samples per class. Furthermore, the meteor and non-meteor events are mostly from camera stations in Germany and the USA, so some possible completely different non-meteor events might not have been observed. 

\textbf{Outliers.} Extremely short and extremely long, bright or large events are very interesting for researchers, so we included as much of them as we were able to. 

\textbf{Measurement bias.} 
Failing measurement methods or devices cause measurement bias. To counter that bias, we used standardized, high-quality cameras with calibrated colors and lenses. Models trained on material collected with other cameras that are equally calibrated should therefore have equal performance. 
The events occurring in the sky are, in most cases, non-meteor events. Since Allsky7 uses simple rules to detect moving events in the sky, it is not entirely assured that every meteor is detected. The camera hardware could cause this uncertainty, for example, by a too high velocity or a too faint appearance, but also by the software by code errors or inappropriate detection rules. We empirically developed the detection code to maximize the number of events and therefore conclude that the number of event types not represented by our data systematically is low. An analysis of the performance of our labeling experts follows underneath. 

\textbf{Observer bias.} This bias occurs when the experimenter unintentionally alters the results of the experiment. We performed our experiments on random samples and performed cross-validation to counter that bias. We also did not influence the labels or the selection process of the samples. 

\textbf{Exclusion bias.} This bias occurs if the experimenter excludes essential information due to its problematic appearance. We did not exclude any sample and tried to choose more challenging samples intentionally. 

\textbf{Experimentation Bias:}
Although we presented various methods and proof of concept for different experiments, it is not assured that they are optimal. All presented observations are affected by parameter choices. Nevertheless, the experimentation, code, and data are publicly available for researchers to reproduce and develop better machine learning methods.

\textbf{Labeling Quality.} We analyzed the quality of the samples in the train set and therefore drew a random sample of 200 events from the data set, consisting of 100 data points per class, to determine the quality of the labels. This sample was independently checked for label correctness by three additional experts. An incorrect label was assumed when any of the experts came to this view. The investigation revealed that for $1$ of the $200$ data points, it could not be said with certainty whether the label was correct because, according to expert opinion, these events are difficult to classify even for humans and represent borderline cases. We refer to these as weakly labeled in the following. We found for $2$ of the $200$ events, a wrong label. We calculated the Clopper-Pearson interval \cite{Clopper1934} to determine how many samples in the dataset are weakly labeled. The number of weakly labeled samples shows us the limit of human classification quality. Machine learning algorithms trained on weakly labeled samples can advance the limit, providing human operators best guesses for challenging samples.
In this dataset, we calculated the $90\%$ confidence interval resulting in a range of $[0.0003,0.0235]$; this means a $90\%$ probability that the correctly labeled proportion of weakly labeled data lies between $0.03\%$ and $2.35\%$. Respectively for the wrong labeled data, the $90\%$ confidence interval results in $[0.0018,0.0311]$. An algorithm that surpasses $97\%$ percent accuracy provides superhuman performance.



%% file: content/conclusions.tex
\section{Conclusions}\label{sec:conclusion}
 Our experiments show high accuracy in Classification and Extrapolation tasks. We observed that unlabeled data that is publicly available was not sufficient to achieve conclusive feature representations and that anomaly detection works not well enough. We, therefore, collected the NightSkyUCP dataset that we made public along with this manuscript to enable researchers to reproduce our experiments and develop their methods on detecting meteors in video data. The dataset consists of 20,000 samples split into 10,000 meteor and 10,000 non-meteor events. A four-step collection process ensures the high quality of the dataset. With this dataset, we showed that the classification efforts of camera operators could be reduced beyond human capabilities.
 Our experiments on anomaly detection show that we can not provide a high accuracy object classification without a negative class. Our observation is conclusive to other observations that show that anomaly detection is difficult, if not impossible \cite{anomaly_review,anomaly_review_IEEE,anomaly_review_PLOS,anomaly_theory}. In our case, we collected and provided negative class samples. Future work will remain to develop an anomaly detection algorithm that can learn what meteors are and how they are separated from anything else. Those algorithms, however, can be evaluated on our dataset. Good candiates for such algorithm are such that introduce prior knowledge \cite{anomaly_theory}.
 Feature-learning tasks gave us insights into the inner structure of the data and found that the sub-classes share similar feature representation but that this is not enough for classification. Therefore, future work is collecting more data belonging to sub-classes.
The extrapolation task helps us match events that we detect in several smaller parts due to obstruction of the trajectory by trees or clouds.
Although the experiments show proof of concept and a deeper insight into the dataset's structure, future work is needed to measure the performance of the presented experimental setups relatively.

%% file: content/ack.tex
\section{Acknowledgement}
The work presented here is based on data of the AllSky7 camera network\footnote{\url{https://allsky7.net/}}. The authors thank the network operators for making their data available.
The copyright of the videos lies with the camera operators.
The camera operators from which data was taken are: Markus Kempf, Bischbrunn Observatory, Hofheim Observatory, Holzkirchen Observatory, Sirko Molau, Jörg Strunk, Kirchheim Observatory, Drebach Observatory, Martin Fiedler and Mike Hankey.
Furthermore, the authors would like to thank Mario Ennes, who has verified the correct labelling of the data. The authors acknowledge financial support by the Carl Zeiss Foundation’s Grant: DeepTurb.

\section{Data availability}
The data underlying this article is available to the reviewers at \hyperlink{https://figshare.com/s/0c88dc958350e32020d7}{figshare} and will be published under a \hyperlink{https://creativecommons.org/licenses/by-nc-sa/4.0/}{CC BY-NC-SA} license once accepted. A datasheet \cite{gebru2018datasheets} summarizing NightSkyUCP along with detailed documentation can be found in the Appendix~A.

%% file: content/ref.tex
\bibliographystyle{agsm}
\bibliography{bibliography.bib}

%% file: content/datasheet.tex
\section{Datasheet}\label{app:datasheet}
\subsection{Motivation}
\textbf{Introduction.} \\
Our goal for this dataset is to stimulate machine learning scientists to design systems that can classify and detect meteors on video data cluttered with objects such as satellites, airplanes, flares, animals, and plants. Under these circumstances, available methods can not accurately provide conclusive results that are needed to gain a seamless overview over the night sky. Consequently, the operation of meteor detection camera systems becomes expensive. However, other available methods, like radar or satellite-based detection, are more expensive and often miss smaller objects. Therefore we made this dataset to reduce the effort of voluntary camera operators to increase the scientific value of wide-area-sky-monitoring.\\
\textbf{Creator.}\\
dAIsy department of the Technical University Ilmenau, Sonneberg Observatory and AllSky7 Network \\
\subsection{Composition}
\textbf{Data description and types.} \\
The dataset consists of:
\begin{itemize}
    \item H264 encoded videos of phenomena inside MPEG-4 container files
    \item absolute and relative positions of the phenomenon in the video and class information in a CSV file
    \item corrected sum images of the video files as jpeg image-files
    \item sequences of 32x32 pixel crops as a NumPy array stack, stored in python pickle format
\end{itemize}  
\textbf{Data size.}\\
The dataset contains 20000 samples belonging to two balanced classeds. 300 meteor and non-meteor samples were additionally validated by five experts and marked as test samples. A sample is discarded if anyone expert voted not entirely sure. Therefore the test set consists of 297 non-meteor and 277 meteor samples.\\
\textbf{Data-context}\\
The NightSkyUCP Dataset is a subset of the data collected and validated of the ongoing AllSky7 sky observation. Samples initially auto-classified as meteor are selected and divided into a meteor and non-meteor classes by experts. It, therefore, contains only those phenomena that are considered difficult to classify by classical image processing techniques. The non-meteor classes are furthermore divided into sub-classes.\\
\textbf{Sample type.}\\
Each sample of an observed phenomenon consists of a video file, metadata such as class, subclass, subset, position, and velocity, and an extracted image stack.\\
\textbf{Sample classes.}\\
All samples are labeled either \emph{meteor} or \emph{non-meteor}. A subset of 116 non-meteor events is further divided into the subclasses \textit{clouds}, \textit{birds}, \textit{planes}, \textit{trees}, \textit{rain} and \textit{light flashes}.\\
\textbf{Known missing sample data}\\
All information needed for the tasks is stored in the dataset. There is no information missing.\\
\textbf{Description of samples' relations.}\\
Samples are stored as sum-images, videos, image-stacks, and metadata-CSV separately. All relations are made explicit.\\
\textbf{Split policy.}\\
We provide a test split that is reviewed by five experts to provide a baseline for training an algorithm. We recommend an 80:20 or 50:50 split for conclusive results.\\
\textbf{Known errors and limitations.}\\
The quality was by five experts' opinions evaluated on 200 samples (100 of each class). Two incorrect samples were found as by two experts contradicted the initial classification. One sample was marked as weakly labeled because experts felt unsure. A Clopper-Pearson test found a 90\% confidence interval of [0.0003, 0.0235] for weakly labeled data and a 90\% confidence interval of [0.0018, 0.0311] for wrongly classified data. \\
\textbf{Needed software to read the data.}\\
The data is hosted by figshare. To read the stack data, the python module pickle is needed in addition to the JPEG and video H264 decoders. \\
\textbf{Confidentiality.}\\
No, the dataset does not contain any confidential information.\\
\textbf{Trauma triggers and trigger warnings.}\\
No, there is no offensive or threatening data included in the dataset. Large fireballs could cause anxiety but are extremely rare, and it is extremely uncertain about getting hit by a meteorite.\\
\textbf{Recording circumstances.}\\
The data was directly recorded as video data at different stations of the AllSky7-Camera-Network. The data were filtered by the camera software and labeled afterward by two experts.\\
\textbf{Data sampling techniques.}\\
The prelabeled by AllSky7 camera software labeled all data samples as meteors in the first place. Afterward, the labels were manually checked by experts \footnote{camera operators and members of the Astronomiemuseum e.V. Sonneberg} and labeled into the meteor and the meteor-like non-meteor class. Afterward, we extracted sequences of image-stacks, sequences metadata, and sum-images from the videos.\\
\textbf{Data sampling strategy.}\\
We sampled non-meteor data that was labeled by the camera software as a meteor from 2019 to 2021. In this period, we were able to sample 10000 events. Afterward, we probabilistically sampled 10000 meteor events also recorded in this period.\\
\textbf{Data sampling worker compensation policy.}\\
All samples were voluntarily labeled by camera operators and experts of the Astronomiemuseum e.V. Sonneberg. The video data was saved automatically and was retrieved by a master's student. \\
\textbf{Data sampling time frame.}\\
All samples are recorded between February 2019 and June 2021.\\
\textbf{Ethical concerns.}\\
There was no ethical review needed.

\subsection{Preprocessing/cleaning/labeling}

 \textbf{Data preprocessing.}\\
The raw video data was pre-labeled by the AllSky7 camera software and reviewed by the corresponding camera operator. For each video following preprocessing steps were carried out for each video: subtraction of the first frame (removing steady objects), the center of gravity detection of the lightest spot, extracting a sequence as 32x32 pixel stack around the lightest spot, extraction of the metadata such as the position of the lightest spot and velocity, and calculation of the sum-image.
 \textbf{Raw data location.}\\
The raw videos are provided along with the dataset.\\
 \textbf{Data preprocessing software.}\\
 The preprocessing event detection software is available at \url{https://github.com/mikehankey/amscams}. Preprocessing scrips and the reproduction package are contents of the data package.

\subsection{Uses}
 \textbf{Data uses.}\\
 The classification and clustering models we describe in the main paper are currently used recommend labels to the camera operators.\\
 \textbf{Dataset landing page.}\\
 No, there is no such repository.\\
 \textbf{Possible uses.}\\
 Other uses could be regression or anomaly detection since our non-meteor class can surely be divided into several sub-classes.\\
 \textbf{Possible harms.}\\
There are no undesirable harms we can think of. But users should know that the video data is collected with specific camera systems, lenses, and sensors. Therefore it is not ensured that the results are valid in a general case.\\
 \textbf{Excluded purposes.}\\
 None.

\subsection{Distribution}
 \textbf{Data accessibility.}\\
The dataset will be hosted on Figshare and will be accessible to anyone. \\
\textbf{Data identifier.}\\
We will provide information on how to use the data as contents of the data package. The dataset will receive a DOI once published.\\
\textbf{Publication date.}\\
The dataset will be distributed in late March 2022.\\
\textbf{Data license.}\\
The dataset will be free for contribution and non-commercial use under the creative commons license: CC BY-NC-SA\\
\textbf{Restrictions.}\\
There will be no restrictions for third parties.\\
\textbf{Regulatory Restrictions.}\\
None.

\subsection{Maintenance}

\textbf{Data owners' institution.}\\
The dataset is maintained by the department dAIsy of the TU Ilmenau, the Astronomiemuseum e.V. Sonneberg and Rabea Sennlaub \\
\textbf{Data owner.}\\
The owner can be contacted through \textit{Rabea@Sennlaub.de} or \textit{Martin.Hofmann@tu-ilmenau.de}\\
\textbf{Erratum.}\\
We do not provide an erratum.\\
\textbf{Data corrections.}\\
Updates will be provided when new data is collected, or more sub-classes are available. \\
\textbf{Versioning.}\\
All versions of the dataset will be hosted and maintained in parallel until they are outdated.\\
\textbf{Contribution policy.}\\
We welcome experts to help the community and us with knowledge and new techniques. Finally, we welcome experts to label more data in more sub-classes.